\documentclass[nohyper, pdftex]{JHEP3}
\usepackage{amsmath,amssymb}
\usepackage{afterpage}
\usepackage{graphicx}
\usepackage{cite}

\makeatletter
\DeclareRobustCommand\recite[1]{\begingroup\@fileswfalse\cite{#1}\endgroup}
\makeatother

\newcommand{\PAGEFIGURE}[1]{\FIGURE[!p]{#1}\afterpage\clearpage}

\newcommand{\Dmq}{\Delta m^2}
\newcommand{\Eps}{\varepsilon}
\newcommand{\diag}{\mathop{\mathrm{diag}}}

\title{Testing matter effects in propagation of atmospheric and
  long-baseline neutrinos}

\author{M.~C.~Gonzalez-Garcia\\
  C.N.~Yang Institute for Theoretical Physics, State University of New
  York at Stony Brook, Stony Brook, NY 11794-3840, USA,\\
  \textrm{and:} Instituci\'o Catalana de Recerca i Estudis
  Avan\c{c}ats (ICREA), Departament d'Estructura i Constituents de la
  Mat\`eria and Institut de Ciencies del Cosmos, Universitat de
  Barcelona, Diagonal 647, E-08028 Barcelona,
  Spain\\
  E-mail:~\email{concha@insti.physics.sunysb.edu}}

\author{Michele Maltoni\\
  Instituto de F\'{\i}sica Te\'orica UAM/CSIC, Calle de Nicol\'as
  Cabrera 13--15, Universidad Aut\'onoma de Madrid, Cantoblanco,
  E-28049 Madrid, Spain\\
  E-mail:~\email{michele.maltoni@csic.es}}

\author{Jordi Salvado\\
  Departament d'Estructura i Constituents de la Mat\`eria and Institut
  de Ciencies del Cosmos, Universitat de Barcelona, Diagonal 647,
  E-08028 Barcelona, Spain\\
  E-mail:~\email{jsalvado@ecm.ub.es}}

\abstract{We quantify our current knowledge of the size and flavor
  structure of the matter effects in the evolution of atmospheric and
  long-baseline neutrinos based solely on the analysis of the
  corresponding neutrino data.  To this aim we generalize the matter
  potential of the Standard Model by rescaling its strength, rotating it
  away from the $ee$ sector, and rephasing it with respect to the
  vacuum term. This phenomenological parametrization can be easily
  translated in terms of non-standard neutrino interactions in matter.
  We show that in the most general case, the strength of the potential
  cannot be determined solely by atmospheric and long-baseline
  data. However its flavor composition is very much constrained and
  the present determination of the neutrino masses and mixing is
  robust under its presence.  We also present an update of the
  constraints arising from this analysis in the particular case in
  which no potential is present in the $e\mu$ and $e\tau$ sectors.
  Finally we quantify to what degree in this scenario it is possible
  to alleviate the tension between the oscillation results for
  neutrinos and antineutrinos in the MINOS experiment and show the
  relevance of the high energy part of the spectrum measured at
  MINOS.}


\preprint{YITP-SB-11-07\\
  IFT-UAM/CSIC-11-07}

\begin{document}
\section{Introduction}

It is now an established fact that neutrinos are massive and leptonic
flavors are not symmetries of Nature~\cite{Pontecorvo:1967fh,
  Gribov:1968kq}.  In the last decade this picture has become fully
proved thanks to the upcoming of a set of precise experiments. In
particular, the results obtained with solar and atmospheric neutrinos
have been confirmed in experiments using terrestrial beams of
neutrinos produced in nuclear reactors and accelerators
facilities~\cite{GonzalezGarcia:2007ib}.
The minimum joint description of all the neutrino data requires mixing
among all the three known neutrinos ($\nu_e$, $\nu_\mu$, $\nu_\tau$),
which can be expressed as quantum superpositions of three massive
states $\nu_i$ ($i=1,2,3$) with masses $m_i$.  This implies the
presence of a leptonic mixing matrix in the weak charged current
interactions~\cite{Maki:1962mu, Kobayashi:1973fv} which can be
parametrized as:
\begin{equation}
  \label{eq:matrix}
  U_\text{vac} =
  \begin{pmatrix}
    c_{12} c_{13}
    & s_{12} c_{13}
    & s_{13} e^{-i\delta_\text{CP}}
    \\
    - s_{12} c_{23} - c_{12} s_{13} s_{23} e^{i\delta_\text{CP}}
    & \hphantom{+} c_{12} c_{23} - s_{12} s_{13} s_{23} e^{i\delta_\text{CP}}
    & c_{13} s_{23} \hspace*{5.5mm}
    \\
    \hphantom{+} s_{12} s_{23} - c_{12} s_{13} c_{23} e^{i\delta_\text{CP}}
    & - c_{12} s_{23} - s_{12} s_{13} c_{23} e^{i\delta_\text{CP}}
    & c_{13} c_{23} \hspace*{5.5mm}
  \end{pmatrix},
\end{equation}
where $c_{ij} \equiv \cos\theta_{ij}$ and $s_{ij} \equiv
\sin\theta_{ij}$.  In addition to the Dirac-type phase
$\delta_\text{CP}$, analogous to that of the quark sector, there are
two physical phases associated to the Majorana character of neutrinos,
which are not relevant for neutrino oscillations~\cite{Bilenky:1980cx,
  Langacker:1986jv} and which are therefore omitted in the neutrino
oscillation analysis.

In the simplest quantum-mechanical picture, flavor oscillations are
generated by the kinematical Hamiltonian for this ensemble
$H_\text{vac}$ which in the flavor basis $(\nu_e, \nu_\mu, \nu_\tau)$
reads
\begin{equation}
  \label{eq:hvac}
  H_\text{vac} = U_\text{vac} D_\text{vac} U_\text{vac}^\dagger
  \quad\text{with}\quad
  D_\text{vac} = \frac{1}{2E_\nu} \diag(0, \Dmq_{21}, \Dmq_{31})
\end{equation}
The quantities $\Dmq_{21}$, $|\Dmq_{31}|$, $\theta_{12}$, and
$\theta_{23}$ are relatively well determined by the analysis of solar,
atmospheric, reactor and long-baseline (LBL) experiments, while only
an upper bound is derived for the mixing angle $\theta_{13}$ and
barely nothing is known on the CP phase $\delta_\text{CP}$ and on the
sign of $\Dmq_{31}$~\cite{Schwetz:2011qt, GonzalezGarcia:2010er,
  Fogli:2009zza, Maltoni:2008ka}.
Given the observed hierarchy between the solar and atmospheric
mass-squared splittings there are two possible non-equivalent
orderings for the mass eigenvalues, which are conventionally chosen as
\begin{align}
  \label{eq:normal}
  \Dmq_{21} &\ll (\Dmq_{32} \simeq \Dmq_{31})
  \text{ with } (\Dmq_{31} > 0) \,;
  \\
  \label{eq:inverted}
  \Dmq_{21} &\ll |\Dmq_{31} \simeq \Dmq_{32}|
  \text{ with } (\Dmq_{31} < 0) \,.
\end{align}
As it is customary we refer to the first option,
Eq.~\eqref{eq:normal}, as the \emph{normal} scheme, and to the second
one, Eq.~\eqref{eq:inverted}, as the \emph{inverted} scheme; in this
form they correspond to the two possible choices of the sign of
$\Dmq_{31}$.\footnote{In this convention the angles $\theta_{ij}$ can
  be taken without loss of generality to lie in the first quadrant,
  $\theta_{ij} \in [0, \pi/2]$, and the CP phase $\delta_\text{CP} \in
  [0, 2\pi]$.}

The flavor evolution of this neutrino ensemble is also affected by the
difference in the matter potential induced by neutrino-matter
interactions when it propagates in a background of sufficiently dense
matter through the so-called Mikheev-Smirnov-Wolfenstein (MSW)
mechanism~\cite{Wolfenstein:1977ue, Mikheev:1986gs}.  Within the
context of the Standard Model (SM) of particle interactions, this
effect is fully determined and leads to a matter potential which for
neutral matter is proportional to the number density of electrons in
the background $N_e(r)$, $V=\sqrt{2} G_F N_e(r)$, and which only
affects electron neutrinos. The evolution of the ensemble is then
determined by the Hamiltonian $ H^\nu = H_\text{vac} +
H^\text{SM}_\text{mat}$, with $H^\text{SM}_\text{mat} = \sqrt{2} G_F
N_e(r) \diag(1, 0, 0)$.  The presence and magnitude of this potential
in the propagation of solar neutrinos can be tested in solar neutrino
experiments (and in combination with KamLAND), and, as pointed out in
Ref.~\cite{Fogli:2002hb}, it agrees well with the SM
prediction. Conversely the effects associated with non-standard forms
of the matter potential in solar neutrino propagation have been
studied and constrained by the analysis of solar and KamLAND
data~\cite{Roulet:1991sm, Guzzo:1991hi, Barger:1991ae, Fogli:1993xv,
  Bergmann:1997mr, Bergmann:2000gp, Guzzo:2000kx, Friedland:2004pp,
  Escrihuela:2009up, Bolanos:2008km, Minakata:2010be, Palazzo:2011vg}.

Matter background effects also affect the evolution of atmospheric and
LBL neutrinos when traveling in the Earth.  However standard matter
effects for atmospheric and LBL neutrino oscillations are suppressed
when compared to solar neutrinos.  This is so because atmospheric and
LBL neutrinos are dominantly $\nu_\mu$'s at the source, the Earth
matter density is smaller, and the characteristic length of the
neutrino path in matter is shorter. Nevertheless the presence of
non-standard matter effects are known to be relevant when discussing
the precise determination of the oscillation parameters in the present
and near future atmospheric and LBL experiments~\cite{Grossman:1995wx,
  GonzalezGarcia:2001mp, Gago:2001xg, Fornengo:2001pm, Huber:2001zw,
  Ota:2001pw, Huber:2002bi, Campanelli:2002cc, Ota:2002na,
  GonzalezGarcia:2004wg, Friedland:2004ah, Friedland:2005vy,
  Blennow:2005qj, Kitazawa:2006iq, Friedland:2006pi, Blennow:2007pu,
  Kopp:2007mi, Kopp:2007ne, Ribeiro:2007ud, Bandyopadhyay:2007kx,
  Ribeiro:2007jq, EstebanPretel:2008qi, Blennow:2008ym, Kopp:2008ds,
  Ohlsson:2008gx, Palazzo:2009rb}.

In this article we address our current knowledge of the size and
flavor structure of the matter background effects in the evolution of
atmospheric and LBL neutrinos based exclusively on the analysis the
present data.  In order to do so we introduce in Sec.~\ref{sec:forma}
a generalized phenomenological parametrization of the matter potential
allowing for rescaling of its strength from the SM prediction,
rotation from the $ee$ sector, and rephasing with respect to
$H_\text{vac}$.  We also discuss the connection with the matter
potential induced by non-standard neutrino interactions (NSI) in
matter which provide a well-known theoretical framework for this
phenomenological parametrization.  In Sec.~\ref{sec:6par} we present
the results of our determination of this generalized matter potential
from the analysis of atmospheric, LBL and CHOOZ experiments.  In
particular we will show that in the most general case, the strength of
the potential cannot be determined by these data. However its flavor
composition is very much constrained.  We will also show that the
present determination of the neutrino masses and mixing is robust even
in this generalized scenario. In Sec.~\ref{sec:4par} we will discuss
the particular case in which no potential is present in the $e\mu$ and
$e\tau$ sector. In this case, a strong bound on the strength of the
potential arises from the analysis of atmospheric and LBL neutrinos.
We will update the results of Ref.~\cite{GonzalezGarcia:2007ib,
  GonzalezGarcia:2004wg} and we will revisit the claims that within
this scenario it is possible to alleviate the tension between the
oscillation results for neutrinos and antineutrinos in the MINOS
experiment.  In Sec.~\ref{sec:conclu} we will draw our conclusions.
Some technical details on the choice of parametrization of the matter
potential are given in two Appendices.

\section{Formalism}
\label{sec:forma}

In the three-flavor oscillation picture, the neutrino (and
antineutrino) evolution equation reads:
\begin{equation}
  i\frac{d}{dx}
  \begin{pmatrix}
    \nu_e\\
    \nu_\mu\\
    \nu_\tau
  \end{pmatrix}
  = H^\nu
  \begin{pmatrix}
    \nu_e\\
    \nu_\mu\\
    \nu_\tau
  \end{pmatrix}
\end{equation}
where $x$ is the coordinate along the neutrino trajectory and the
Hamiltonian for neutrinos and antineutrinos is:
\begin{equation}
  H^\nu = H_\text{vac} + H_\text{mat}
  \quad\text{and}\quad
  H^{\bar\nu} = ( H_\text{vac} - H_\text{mat} )^* \: ,
\end{equation}
with $ H_\text{vac}$ given in Eq.~\eqref{eq:hvac}.
In the Standard Model $H_\text{mat}$ is fully determined both in its
strength and flavor structure to be $H^\text{SM}_\text{mat} =\sqrt{2}
G_F N_e(r) \diag(1, 0, 0)$.  In this work we generalize the form of
the matter potential to be
\begin{equation}
  \label{eq:hmatpar}
  H_\text{mat} = Q_\text{rel} U_\text{mat} D_\text{mat}
  U_\text{mat}^\dagger Q_\text{rel}^\dagger
  \quad\text{with}\quad
  \left\lbrace
  \begin{aligned}
    Q_\text{rel} &= \diag\left(
    e^{i\alpha_1}, e^{i\alpha_2}, e^{-i\alpha_1 -i\alpha_2} \right),
    \\
    U_\text{mat} &= R_{12}(\varphi_{12}) R_{13}(\varphi_{13}) \,,
    \\
    D_\text{mat} &= \sqrt{2} G_F N_e(r) \diag(\Eps, 0, 0)
  \end{aligned}\right.
\end{equation}
where we denote by $R_{ij}(\varphi_{ij})$ a rotation of angle
$\varphi_{ij}$ in the $ij$ plane.
In this parametrization $\Eps$ represents a rescaling of the matter
potential strength, $\varphi_{12}$ allows for projection of the
potential into the $\nu_\mu$ flavor and $\varphi_{13}$ allows for its
projection into the $\nu_\tau$ flavor.  The two additional phases
$\alpha_1$ and $\alpha_2$ included in $Q_\text{rel}$ are \emph{not} a
feature of neutrino-matter interactions, but rather a relative feature
of the vacuum and matter term: they would become unphysical if any of
the two terms weren't there.  As described in Appendix~\ref{sec:app2}
one can always take $0 \le \theta_{ij} \le \pi/2$ and $\Dmq_{31} \ge
0$ and the neutrino mass hierarchy is accounted by the sign of $\Eps$:
positive for normal ordering, negative for inverted ordering.  For the
case of \emph{real} potential ($\alpha_1 = \alpha_2 = 0$) one must
consider $-\pi/2 \le \varphi_{ij} \le \pi/2$, whereas for complex NSI
(free $\alpha_1$ and $\alpha_2$, $0\le \alpha_i\le 2\pi$) it is enough
to assume $0 \le \varphi_{ij} \le \pi/2$.  In this parametrization the
standard $3\nu$ oscillations in matter are recovered in the limit
$\Eps = \pm 1$ and $\varphi_{12} = \varphi_{13} = 0$, where $\Eps =
+1$ corresponds to the normal ordering and $\Eps = -1$ corresponds to
the inverted ordering.

As explained in Appendix~\ref{sec:app1} (see Eq.~\eqref{eq:hmatgen}
and below) Eq.~\eqref{eq:hmatpar} is not the most general
parametrization for the matter potential, since it has built-in the
assumption that two of its eigenvalues are equal.  In
Ref.~\cite{Friedland:2004ah} it was shown that strong cancellations in
the oscillation of atmospheric neutrinos occur in this case.
Consequently despite our phenomenological parametrization in
Eq.~\eqref{eq:hmatpar} represents only a subspace of the most general
parameter space for the matter potential, it is precisely in this
subspace where the weakest constraints can be placed.

For the analysis of atmospheric, LBL and CHOOZ data, one can make the
additional simplifying assumption of setting $\Dmq_{21} = 0$, hence
neglecting the solar splitting.  In this limit the $\theta_{12}$ angle
and the $\delta_\text{CP}$ phase become unphysical, even in the
presence of the generalized $H_\text{mat}$ introduced above. Thus
altogether the relevant flavor transition probabilities for
atmospheric, LBL and CHOOZ neutrinos depend on eight parameters:
($\Dmq_{31}$, $\theta_{13}$, $\theta_{23}$) for the vacuum part,
($\Eps$, $\varphi_{12}$, $\varphi_{13}$) for the matter part, and
($\alpha_1$, $\alpha_2$) as relative phases.

We notice that within this parametrization and neglecting $\Dmq_{21}$
both the vacuum and the matter part of the Hamiltonian have each two
degenerate eigenvalues.  In Ref.~\cite{Blennow:2008eb} it was noted
that when this occurs the evolution equation can be reduced to an
effective two-flavor problem in terms of an intermediate basis. Such
effective $2\times 2$ evolution Hamiltonian depends only on 3
parameters: $\Dmq_{31}$, $\Eps$, and an angle which is a given
function $\theta_{13}$, $\theta_{23}$, $\varphi_{12}$ and
$\varphi_{13}$ (see Appendix~\ref{sec:app1} for details).  Therefore
we can solve the evolution in this reduced parameter space and later
on project back to the flavor basis.  Technically this makes it
possible to perform the analysis of the atmospheric, LBL and CHOOZ
data in the full eight-dimensional parameter space.

A theoretical framework for our proposed parametrization of the matter
potential is provided by NSI affecting neutrino interactions in the
Earth matter.  They can be described by effective four-fermion
operators of the form
\begin{equation}
  \label{eq:def}
  \mathcal{L}_\text{NSI} =
  - 2\sqrt{2} G_F \Eps^{fP}_{\alpha\beta}
  (\overline{\nu_{\alpha}} \gamma^\mu \nu_{\beta})
  (\overline{f}\gamma_\mu P f)\, ,
\end{equation}
where $f$ is a charged fermion, $P=(L,R)$ and
$\Eps^{fP}_{\alpha\beta}$ are dimensionless parameters encoding the
deviation from standard interactions.
Ordinary matter is composed by electrons ($e$), up-quarks ($u$) and
down-quark ($d$), and in principle non-standard neutrino-matter
interactions can involve any of these particles with different
strength. In practice, however, the proton/neutron ratio found in
matter is reasonably constant all over the Earth, and pretty close to
$1$. This implies that neutrino oscillations are only sensitive to the
\emph{sum} of these interactions, weighted with the relative abundance
of each particle.  Furthermore NSI enter in neutrino propagation only
through the vector couplings $\Eps^f_{\alpha\beta} =
\Eps^{fL}_{\alpha\beta} + \Eps^{fR}_{\alpha\beta}$. We can therefore
define the relevant combination of NSI parameters entering into the
neutrino propagation in the Earth as:
\begin{equation}
  \Eps_{\alpha\beta} \equiv
  \sum_{f=e,u,d} \left< \frac{N_f}{N_e} \right> \Eps_{\alpha\beta}^f
  \approx \Eps_{\alpha\beta}^e
  + 3 \Eps_{\alpha\beta}^u
  + 3 \Eps_{\alpha\beta}^d
\end{equation}
where $N_f$ is the number density of the fermion $f$. The
corresponding matter Hamiltonian reads:
\begin{equation}
  \label{eq:hmatepsi}
  H_\text{mat} = \sqrt{2} G_F N_e(r)
  \begin{pmatrix}
    1 + \Eps_{ee} & \Eps_{e\mu} & \Eps_{e\tau}
    \\
    \Eps_{e\mu}^* & \Eps_{\mu\mu} & \Eps_{\mu\tau}
    \\
    \Eps_{e\tau}^* & \Eps_{\mu\tau}^* & \Eps_{\tau\tau}
  \end{pmatrix} \,.
\end{equation}
which includes both the standard (accounted by the ``$+1$'' term in
the $ee$ entry) and the non-standard interactions (accounted by the
$\Eps_{\alpha\beta}$ terms).

The Hermitian matrix $H_\text{mat}$ depends on 8 physical parameters.
Oscillation experiments are sensitive to the matter potential up to an
overall multiple of the identity, thus we can measure two differences
of the $\Eps_{\alpha\alpha}$ parameters (which must be real) as well
as $\Eps_{e\mu}$, $\Eps_{e\tau}$ and $\Eps_{\mu\tau}$ (which, in
general, can be complex).  These 8 parameters are reduced to the 5 in
Eq.~\eqref{eq:hmatpar} under the condition that two eigenvalues of
$H_\text{mat}$ are equal, which eliminates one real parameter, one
angle and one phase leading to:
\begin{equation}
  \label{eq:epsis}
  \begin{aligned}
    \Eps_{ee} - \Eps_{\mu\mu}
    &= \hphantom{-} \Eps \, (\cos^2\varphi_{12} - \sin^2\varphi_{12})
    \cos^2\varphi_{13} - 1 \,,
    \\
    \Eps_{\tau\tau} - \Eps_{\mu\mu}
    &= \hphantom{-} \Eps \, (\sin^2\varphi_{13}
    - \sin^2\varphi_{12} \, \cos^2\varphi_{13}) \,,
    \\
    \Eps_{e\mu}
    &= -\Eps \, \cos\varphi_{12} \, \sin\varphi_{12} \,
    \cos^2\varphi_{13} \, e^{i(\alpha_1 - \alpha_2)} \,,
    \\
    \Eps_{e\tau}
    &= -\Eps \, \cos\varphi_{12} \, \cos\varphi_{13} \,
    \sin\varphi_{13} \, e^{i(2\alpha_1 + \alpha_2)} \,,
    \\
    \Eps_{\mu\tau}
    &= \hphantom{-} \Eps \, \sin\varphi_{12} \, \cos\varphi_{13} \,
    \sin\varphi_{13} \, e^{i(\alpha_1 + 2\alpha_2)} \,.
  \end{aligned}
\end{equation}
Notice that the term ``${}-1$'' at the end of $\Eps_{ee} -
\Eps_{\mu\mu}$ accounts for the standard matter term.  In terms of the
NSI parameters $\Eps_{\alpha\beta}$ the magnitude of $\Eps$ can be
always expressed as:
\begin{equation}
  \begin{aligned}
    \Eps^2 &= \frac{3}{2} \min_\lambda \Bigg[ \sum_{\alpha\beta}
      \big| \delta_{e\alpha} \delta_{e\beta} + \Eps_{\alpha\beta} -
      \lambda \delta_{\alpha\beta} \big|^2 \Bigg]
    \\
    &= \frac{1}{2} \Bigg[ 3 \sum_{\alpha\beta}
      \big| \delta_{e\alpha} \delta_{e\beta} + \Eps_{\alpha\beta} \big|^2
      - \Big( 1 + \sum_\alpha \Eps_{\alpha\alpha} \Big)^2 \Bigg] \,,
  \end{aligned}
\end{equation}
and its sign can be defined by the relative size of the eigenvalues of
$H_\text{mat}$: $\Eps$ is positive (negative) if the two equal
eigenvalues are smaller (larger) than the third one.

As mentioned above in Ref.~\cite{Friedland:2004ah} it was shown that
strong cancellations (hence weak bounds on the $\Eps$'s) in the
oscillation of atmospheric neutrinos occur if two of the NSI
eigenvalues coincide. This is so because in this limit there is a
two-dimensional subspace where $\nu_\mu \rightarrow \nu_\tau$
oscillations occur as in vacuum.  Consequently, although
Eq.~\eqref{eq:hmatpar} cover only a part of the most general parameter
space for the matter potential induced by NSI, it is the part where
the weakest constraints on NSI can be placed.  Consequently our
conclusions on the allowed values of the NSI parameters are not
expected to be weakened once the ``equal eigenvalues'' approximation
is removed.

We conclude this section by noticing that until recently it was
usually assumed that the bounds from loop contributions to flavor
changing processes induced by the NSI implied $\Eps_{e\mu} \lesssim
10^{-4}$~\cite{Davidson:2003ha}.  If the condition $\Eps_{e\mu} = 0$
is imposed (together with the assumption that two NSI eigenvalues are
equal) one is left with two possible scenarios: (i)
$\Eps_{\mu\tau}=0$, which is a subset of the $e\tau$ sector studied in
Refs.~\cite{Friedland:2004ah, Friedland:2005vy} (who made the
assumption $\Eps_{e\mu} = \Eps_{\mu\tau} = 0$ but left both
eigenvalues free), and (ii) $\Eps_{e\tau} = 0$, which corresponds to
the NSI being dominantly in the $\mu\tau$
sector~\cite{Fornengo:2001pm, GonzalezGarcia:2004wg,
  GonzalezGarcia:2007ib} and is revisited in Sec.~\ref{sec:4par}.
However, in Ref.~\cite{Biggio:2009kv} it was argued that the loop
bound on $\Eps_{e\mu}$ does not hold in general and consequently it
raised the interest of the general analysis here presented.

\section{Results}
\label{sec:6par}

Let us now present the results of a combined analysis of atmospheric,
LBL and CHOOZ experiments in the context of oscillations with the
generalized matter potential in the Earth in Eq.~\eqref{eq:hmatpar}.
It is well known that important constraints on $\nu_e$ flavor
oscillations driven by $\Dmq_{31}$ arises from the negative results on
$\bar\nu_e$ disappearance at short baselines at the CHOOZ reactor
experiment~\cite{Apollonio:1999ae}.  We include in our analysis the
results of CHOOZ under the assumption that, because of the short
baseline, matter effects are irrelevant in this experiment.
For atmospheric neutrinos we include the results of our reanalysis of
the Super-Kamiokande I+II+III phases as presented
in~\cite{Wendell:2010md} and described in
Ref.~\cite{GonzalezGarcia:2010er}; details on our simulation of the
data samples and the statistical analysis can be found in the Appendix
of Ref.~\cite{GonzalezGarcia:2007ib}.
For what concerns LBL accelerator experiments, we combine the results
of the energy spectrum of $\nu_\mu$ events obtained by MINOS after an
exposure to the Fermilab NuMI beam corresponding to a total of $7.25
\times 10^{20}$ protons on target~\cite{Adamson:2011ig} with the data
on antineutrino $\bar\nu_\mu$ disappearance corresponding to $1.71
\times 10^{20}$ protons on target~\cite{Adamson:2011fa}.  We also
account for the recent MINOS results on $\nu_\mu\rightarrow \nu_e$
transitions for which we fit the total event rate of observed $\nu_e$
events above the expected background based on an exposure of $7.01
\times 10^{20}$ protons on target~\cite{Adamson:2010uj}.

\FIGURE[t]{
  \includegraphics[width=0.9\textwidth]{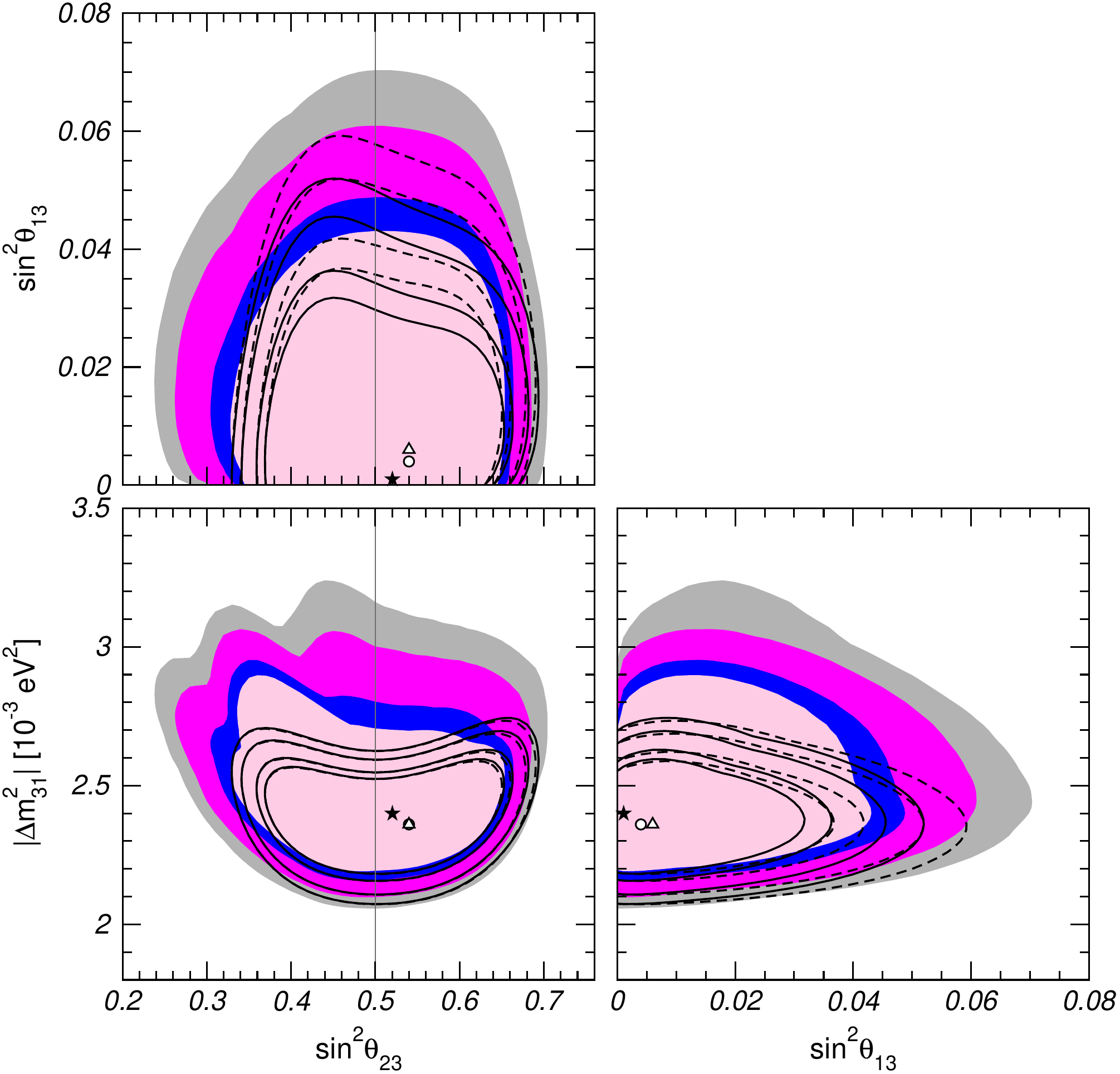}
  \caption{Two-dimensional projections of the allowed regions from the
    global analysis of atmospheric, LBL and CHOOZ data in the
    oscillation parameters $|\Dmq_{31}|$, $\sin^2\theta_{23}$ and
    $\sin^2\theta_{13}$ after marginalizing over $\Eps$,
    $\varphi_{12}$, and $\varphi_{13}$ (full regions).  The best fit
    point is marked with a star.  For the sake of comparison we also
    show in the figure the corresponding regions for the case of
    standard matter potential with normal ordering ($\Eps = +1$,
    $\varphi_{12} = \varphi_{13} = 0$) as unfilled regions with full
    lines with best fit point marked with a circle and inverted
    ordering ($\Eps = -1$, $\varphi_{12} = \varphi_{13} = 0$) as
    unfilled regions with dashed lines with best fit point marked with
    a triangle.  The regions are shown at 90\%, 95\%, 99\% and
    $3\sigma$ CL (2~dof).}
  \label{fig:proj-osc}
}

We present the results of the global analysis in
Figs.~\ref{fig:proj-osc}, \ref{fig:proj-eps}, \ref{fig:range-osc}
and~\ref{fig:chisq}.  For simplicity the results are given for fixed
values of the phases $\alpha_1 = \alpha_2 = 0$. We have verified that
the phases have little impact on the results of this analysis. In
particular, no information on them can be extracted from the data
(\textit{i.e.}, their allowed range is the full $0\le \alpha_i \le
2\pi$ interval) and they have only a minor effect on the determination
of the other six real parameters.

\FIGURE[t]{
  \includegraphics[width=0.9\textwidth]{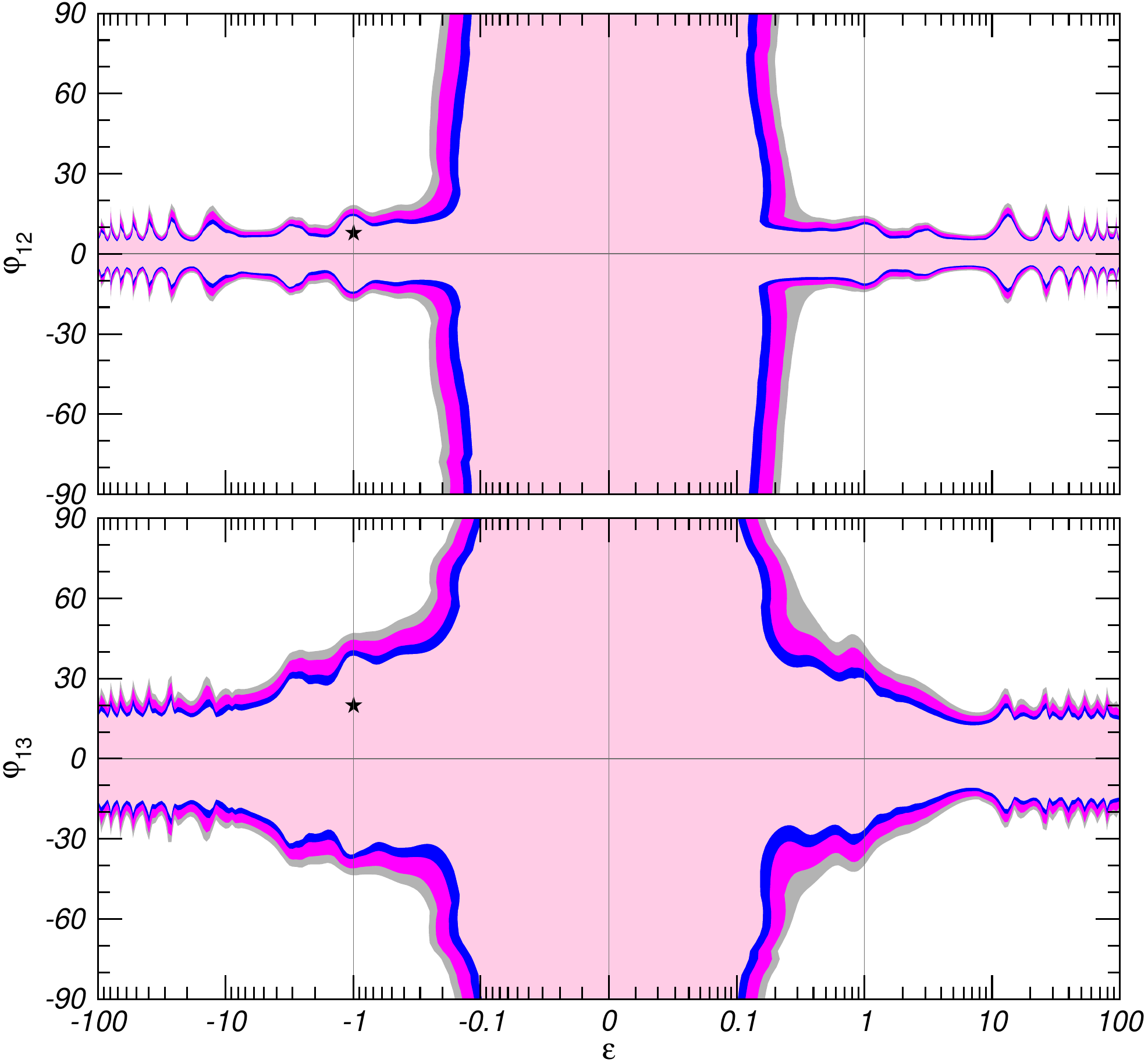}
  \caption{Two-dimensional projection of the of the allowed regions
    from the global analysis of atmospheric, LBL and CHOOZ data in the
    matter potential parameters $\Eps$, $\varphi_{12}$, and
    $\varphi_{13}$ after marginalization with respect to the
    undisplayed parameters.  The regions are shown at 90\%, 95\%, 99\%
    and $3\sigma$ CL (2~dof). The best fit point is marked with a
    star.}
  \label{fig:proj-eps}
}

Fig.~\ref{fig:proj-osc} contains the three two-dimensional projections
of the allowed regions in the oscillation parameters $|\Dmq_{31}|$,
$\sin^2\theta_{23}$ and $\sin^2\theta_{13}$ after marginalizing over
$\Eps$, $\varphi_{12}$, and $\varphi_{13}$. The regions are shown at
90\%, 95\%, 99\% and $3\sigma$ CL (2~dof). Notice that since we have
chosen to absorb the relative sign between the $\Dmq_{31}$ and the
matter potential in $\Eps$ there is only one oscillation region.  For
the sake of comparison we also show in the figure the corresponding
regions for the case of standard matter potential with normal ordering
($\Eps = +1$, $\varphi_{12} = \varphi_{13} = 0$) and inverted ordering
($\Eps = -1$, $\varphi_{12} = \varphi_{13} = 0$).
Fig.~\ref{fig:proj-eps} displays the two-dimensional projections of the
allowed regions in the matter potential parameters $\Eps$,
$\varphi_{12}$ and $\varphi_{13}$ after marginalizing over
$|\Dmq_{31}|$, $\theta_{23}$ and $\theta_{13}$.  The regions are
shown at 90\%, 95\%, 99\% and $3\sigma$ CL (2~dof).
From these figures we can draw the following conclusions on the
determination of the matter potential:
\begin{itemize}
\item No bound on the magnitude of the matter effects, $\Eps$, can be
  derived from the analysis of atmospheric and LBL (and CHOOZ)
  experiments only in this general scenario.  Consequently in order to
  attain bounds on $\Eps$ it is necessary to combine these results
  with those from other neutrino oscillation samples involving $\nu_e$
  oscillations, such as solar and KamLAND experiments. However, a
  bound on $\Eps$ can be derived if a certain flavor structure of the
  matter potential is assumed \emph{a priori}, implying that
  $\varphi_{12}$ and/or $\varphi_{13}$ are larger than a certain
  amount. This is the case, for example, if we assume that no matter
  effects are present in the $e\mu$ and $e\tau$ projections, which
  corresponds to $\varphi_{12} = \pi/2$. We will study this case in
  the next section.

\item Matter potentials with $|\Eps| \gtrsim \mathcal{O}(0.2)$ are
  only allowed as long as their flavor projections out of the $ee$
  entry are severely constrained.  The constraint is stronger on
  rotations over the $\mu$ flavor, $\varphi_{12}$. We also notice that
  for large values of $\Eps$ the precise bounds on $\varphi_{12}$ and
  $\varphi_{13}$ present an oscillatory behavior. We will discuss the
  origin of this beaviour below.
\end{itemize}
Qualitatively these results can be understood as follows.  For
$\sqrt{2} G_F N_e(r) \Eps \gg \Dmq_{31} / (2 E_\nu)$ the neutrinos can
undergo flavor oscillations with two different phases:
\begin{align}
  \label{eq:Dvac}
  \tilde{\Delta}_\text{vac}
  &= \frac{\Dmq_{31} L}{4 E_\nu} \times
  f(\theta_{23},\theta_{13},\varphi_{12},\varphi_{13})
  \\
  \label{eq:Dmat}
  \tilde{\Delta}_\text{mat}
  &= \frac{\sqrt{2} G_F N_e(r) \Eps L}{2}
\end{align}
where $f(\theta_{23}, \theta_{13}, \varphi_{12}, \varphi_{13})$ is a
combination of trigonometric functions of the four angles: for
example, for $\theta_{13}=0$ we get $f = \cos^2\varphi_{12}
\cos^2\varphi_{13} + (\cos\varphi_{13} \sin\varphi_{12}
\cos\theta_{23} - \sin\varphi_{13} \sin\theta_{23})^2$.  $\nu_\mu
\leftrightarrow \nu_\tau$ transitions oscillate dominantly with phase
$\tilde{\Delta}_\text{vac}$: this is the generalization of the results
in Ref.~\cite{Friedland:2004ah, Friedland:2005vy} and it means that in
this scenario $\nu_\mu\rightarrow \nu_\tau$ oscillations have the same
dependence on the neutrino energy and distance as vacuum oscillations,
even for large $\Eps$, which opens the possibility of giving a good
description of atmospheric and LBL results.
In the same regime $\nu_\mu\leftrightarrow \nu_e$ and
$\nu_e\leftrightarrow \nu_\tau$ transitions proceed dominantly via
oscillations of phase $\tilde{\Delta}_\text{mat}$
\begin{align}
  \label{eq:Pem}
  P_{e\mu} &= \cos^2\varphi_{13} \, \sin^2(2\varphi_{12}) \,
  \sin^2\tilde{\Delta}_\text{mat} \,,
  \\
  \label{eq:Pet}
  P_{e\tau} &= \cos^2\varphi_{12} \, \sin^2(2\varphi_{13}) \,
  \sin^2\tilde{\Delta}_\text{mat} \,.
\end{align}
This means that even for $\theta_{13}=0$ atmospheric $\nu_e$'s can
disappear into either $\nu_\mu$ or $\nu_\tau$ and they do so
independently of their energy.  Similarly, some of the atmospheric
muon neutrinos will oscillate into electron neutrinos independently of
their energy. This is in clear conflict with the atmospheric data and
therefore the amplitudes of these oscillations are constrained. This
leads to the strong bounds on $|\varphi_{12}|$ and $|\varphi_{13}|$
observed in Fig.~\ref{fig:proj-eps} for $|\Eps|\gtrsim
\mathcal{O}(0.2)$.  Furthermore, since the atmospheric $\nu_e$ data
mostly constrain the \emph{sum} of $P_{e\mu} + P_{e\tau}$, the bounds
on $\varphi_{12}$ and $\varphi_{13}$ for a given $\Eps$ are strongly
correlated.  Notice also that atmospheric neutrino angular bins
correspond to given \emph{ranges} of $L$, rather than just a specific
value as in the case of MINOS, so for large $\Eps$ the phase
$\tilde{\Delta}_m$ is averaged in $L$ and the atmospheric bounds on
$\varphi_{12}$ and $\varphi_{13}$ become independent of $\Eps$.

Similarly to the atmospheric neutrino data, the MINOS $\nu_\mu$
disappearance spectrum is well explained by the vacuum-like
oscillations $\nu_\mu\rightarrow \nu_\tau$. But it also leads to
further constraints on the contribution from $P_{e\mu}$
in~\eqref{eq:Pem}, since the presence of this term would induce an
energy independent contribution to the $\nu_\mu$ survival probability
which is in conflict with the observations.  This tightens the
atmospheric constraint on $\varphi_{12}$. Since $L$ in MINOS is fixed,
the contribution of $P_{e\mu}$ to the MINOS spectrum has an
oscillatory behavior on $\Eps$ (\textit{i.e.}, the phase
$\tilde{\Delta}_m$ is \emph{not} averaged over $L$ as long as the size
of the detector is small compared to $[G_F N_e(r) \Eps]^{-1}$) which
leads to the ``spike-like'' shape of the allowed region for
$\varphi_{12}$ versus $\Eps$.  Given the strong correlation between
the allowed values of $\varphi_{13}$ and $\varphi_{12}$ imposed by the
atmospheric data, this ``spike-like'' shape is also projected in the
allowed region for $\varphi_{13}$ versus $\Eps$.

\FIGURE[t]{
  \includegraphics[width=0.9\textwidth]{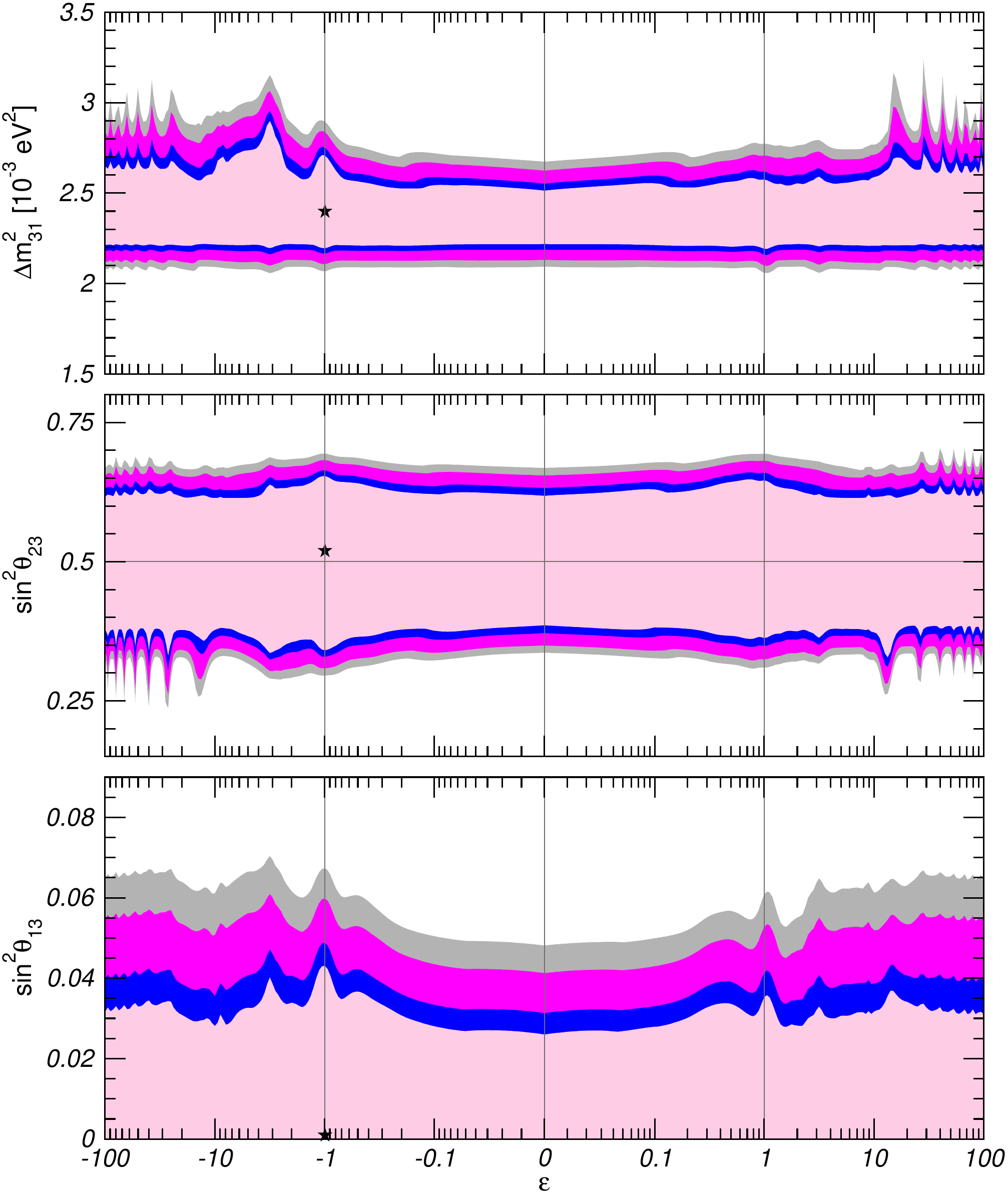}
  \caption{Two-dimensional projection of the of the allowed regions
    from the global analysis of atmospheric, LBL and CHOOZ for the
    oscillation parameters and the $\Eps$ parameter of the matter
    potential magnitude. The regions are shown at 90\%, 95\%, 99\% and
    $3\sigma$ CL (2~dof). The best fit point is marked with a star.}
  \label{fig:range-osc}
}

For what concerns the standard oscillation parameters, our analysis
shows that their present determination is rather robust even in the
presence of a generalized matter potential of the
form~\eqref{eq:hmatpar}. To further illustrate this we show in
Fig.~\ref{fig:range-osc} the allowed regions of the three oscillation
parameters as a function of the matter potential magnitude $\Eps$. As
seen in the figure, the allowed ranges for the oscillation parameters
do not get substantially modified for any value of $\Eps$ (see also
Eq.~\eqref{eq:oscrang}).

\PAGEFIGURE{
  \includegraphics[width=0.96\textwidth]{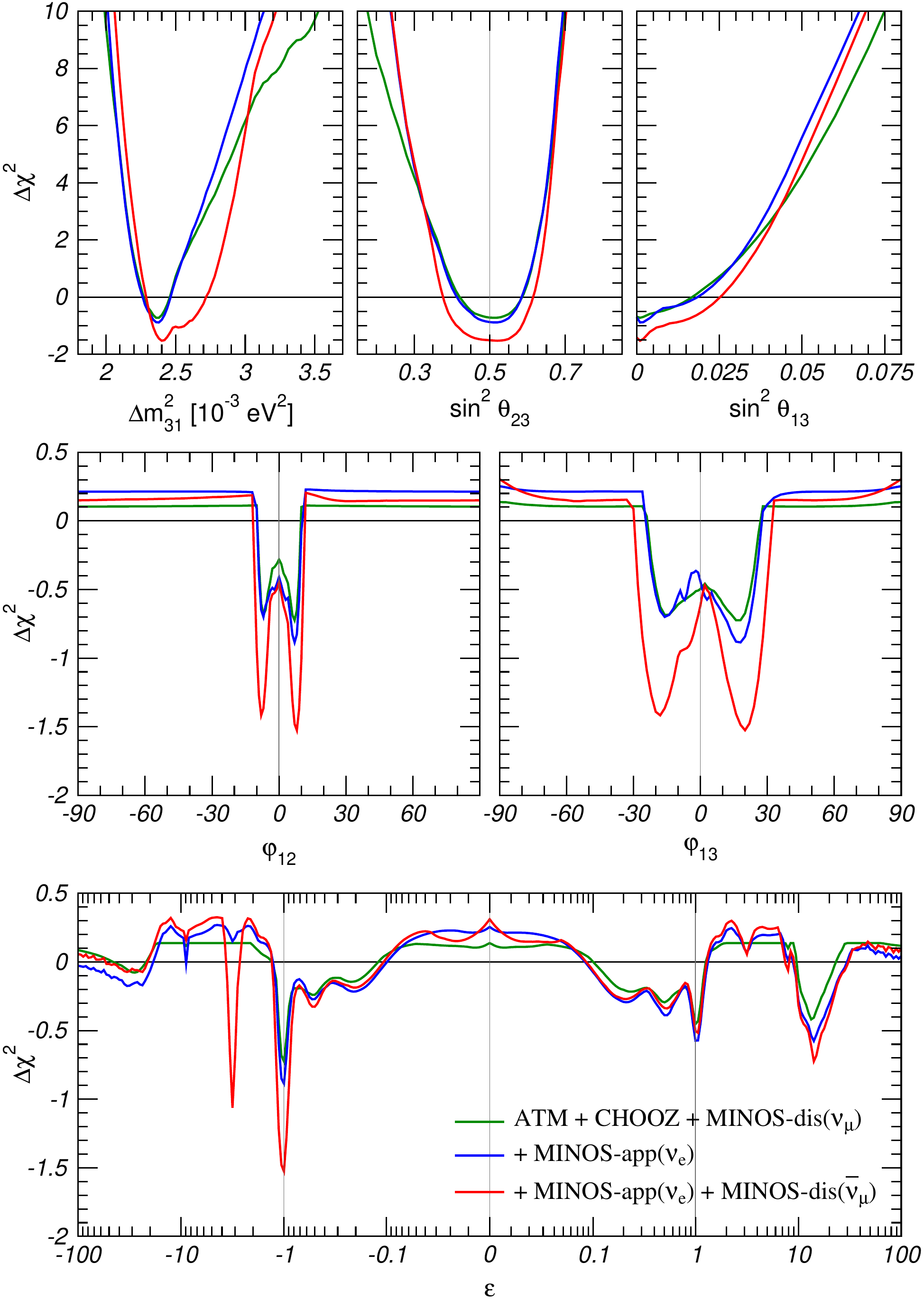}
  \caption{Dependence of the $\Delta\chi^2$ function for different
    combination of observables on the oscillation and matter potential
    parameters. In each panel the $\Delta\chi^2$ function has been
    marginalized with respect to all other parameters. The
    $\Delta\chi^2$ is defined with respect to the minimum for pure
    oscillations which lies in the $\Eps = +1$, $\varphi_{12} =
    \varphi_{13} = 0$ subspace.}
  \label{fig:chisq}
}

The dependence on these results on the data samples included in the
analysis is shown in Fig.~\ref{fig:chisq} where we plot the one
dimensional projections of the $\Delta\chi^2$ for three different
combination of experiments as a function of each of the six
parameters. In each case the $\Delta\chi^2$ is defined with respect to
the minimum obtained for the standard oscillation case which occurs in
the normal ordering ($\Eps = +1$, $\varphi_{12} = \varphi_{13} = 0$).
We read from the figure that for any of the three combination of data
shown, there is no substantial improvement of the fit due to the
inclusion of the extra parameters. $\Delta\chi^2$ is reduced by a
maximum of $1.5$ units in the global analysis. In other words, even
though no constraint on the magnitude of the matter effects can be
placed by the data, the data do not show any need of a non-standard
matter potential.  In particular comparing the results including and
not including the $\bar\nu_\mu$ MINOS results we see that inclusion of
non-standard matter effects does not result into a substantial
improvement on the quality of the combined fit.  From the lower panel
of the figure we see that inclusion of the MINOS antineutrino results
implies the appearance of a new local minimum at $\Eps = -3$. This
local minimum occurs at $\varphi_{13} = -16^\circ$ and $\varphi_{12} =
-7^\circ$ and it is the equivalent of the best solution found in the
partial analysis of Refs.~\cite{Kopp:2010qt, Mann:2010jz,
  Akhmedov:2010vy} to explain the discrepancy of MINOS neutrino and
antineutrino results in the framework of NSI.  However, as seen in the
figure, we find that the ``improvement'' on the quality of the
description associated with this new minimum is about $1$ unit in the
global $\chi^2$.  This is partly due to the constraints from
atmospheric neutrino data on the required values of the NSI and partly
on details associated with the statistical analysis of MINOS in the
presence of NSI, which we will quantify in more detail in the next
section.

As a matter of curiosity we notice that the best fit of the global
analysis corresponds to ``standard'' magnitude matter effects with
inverted mass ordering ($\Eps = -1$), maximal $\theta_{23}$ and
$\theta_{13}=0$, but with non-standard flavor projections
$\varphi_{12} = 8^\circ$ and $\varphi_{13} = 20^\circ$. This behavior
is already present when only atmospheric and CHOOZ data are included
in the analysis and it is driven by the small excess of $\nu_e$ events
in the multi-GeV atmospheric samples. This excess was more significant
in SK-I and was highlighted in Ref.~\cite{Fogli:2008jx} as a possible
``hint'' for non-zero $\theta_{13}$, although as pointed out in
Refs.~\cite{Maltoni:2008ka, GonzalezGarcia:2010er} the statistical
significance of this effect was reduced in SK-II and SK-III
samples. What we find in our analysis is that it is slightly easier to
fit the data with a non-standard flavor structure of the matter
potential and a zero $\theta_{13}$ than with standard matter potential
and a non-vanishing $\theta_{13}$.  Nevertheless the difference,
yielding a $\Delta\chi^2=-1.5$, is not statistically significant.

\PAGEFIGURE{
  \includegraphics[width=\textwidth]{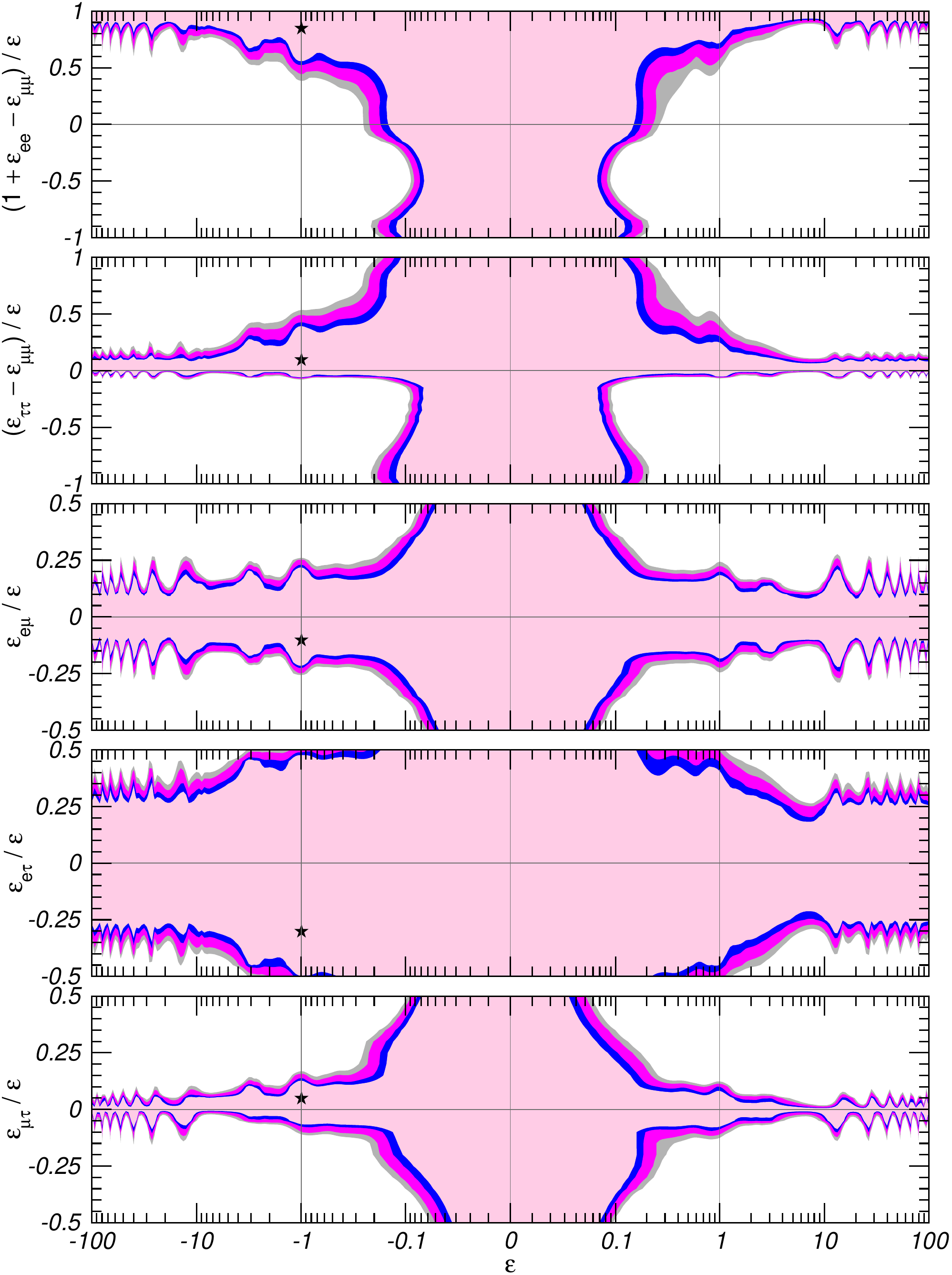}
  \caption{Two-dimensional projection of the of the allowed regions
    from the global analysis of atmospheric, LBL and CHOOZ in terms of
    the NSI parameters $\Eps_{\alpha\beta}$ versus the $\Eps$
    parameter. The regions are shown at 90\%, 95\%, 99\% and $3\sigma$
    CL (2~dof). The best fit point is marked with a star.}
  \label{fig:range-eps}
}

Finally we show in Fig.~\ref{fig:range-eps} the two-dimensional
allowed regions projected in terms of the equivalent NSI parameters
$\Eps_{\alpha\beta}$, see Eq.~\eqref{eq:epsis}. The figure illustrates
that although no absolute bound can be place on the individual
$\Eps_{\alpha\beta}$ parameters, they can only be large if they are
highly correlated among them. In other words, in order to comply with
the observed data in atmospheric, LBL and CHOOZ experiments, a beyond
the SM scenario which could generate such large NSI parameters must
have a very constrained flavor structure which implies, for example,
$\Eps_{\mu\tau} / \Eps \lesssim 0.1$ which, unfortunately, is not the
one required to explain the MINOS $\nu_\mu$ versus $\bar\nu_\mu$
results. We elaborate more on this in the following section.

\section{No matter effects in $e\mu$ and $e\tau$ sectors}
\label{sec:4par}

In this section we study the particular case in which no potential is
present in the $e\mu$ and $e\tau$ sector, or, equivalently, that the
NSI parameters $\Eps_{e\mu} = \Eps_{e\tau} = 0$ are known to be
severely constrained from other sectors.  As described in
Refs.~\cite{Fornengo:2001pm, GonzalezGarcia:2004wg}, in this case a
bound on the strength of the potential arises from the analysis of
atmospheric and LBL neutrinos.
From Eq.~\eqref{eq:epsis} we see that $\Eps_{e\mu} = \Eps_{e\tau} = 0$
corresponds to $\varphi_{12} = \pi/2$. If we further assume
$\theta_{13} = 0$ then the $\nu_e$ decouples from the evolution
equation. Moreover one of the two phases become unphysical so that
the matter part of the Hamiltonian depends on three parameters.  In
Refs.~\cite{GonzalezGarcia:2004wg, GonzalezGarcia:2007ib} we labeled
them $\mathcal{F}$, $\xi$ and $\eta$ and they are related to the
present notation by
\begin{equation}
  \mathcal{F} = \sqrt{|\Eps_{\mu\tau}|^2 +
    \frac{(\Eps_{\mu\mu}-\Eps_{\tau\tau})^2}{4}} = \frac{\Eps}{2} \,,
  \quad
  \xi = \frac{|\Eps_{\mu\tau}|}{\mathcal{F}} = \varphi_{13} \,,
  \quad
  \eta = \arg(\Eps_{\mu\tau}) = \alpha_1+2\alpha_2 \,.
\end{equation}
The relevant oscillation probabilities are given by
\begin{equation}
  \label{eq:prob}
  P_{\nu_\mu \to \nu_\mu} = 1 - P_{\nu_\mu \to \nu_\tau} =
  1 - \sin^2 2\Theta \, \sin^2 \left(
  \frac{\Dmq_{31} L}{4 E_\nu} \, \mathcal{R} \right) \,.
\end{equation}
where
\begin{align}
  \sin^2 2\Theta &= \frac{1}{\mathcal{R}^2} \left(
  \sin^2 2\theta_{23} + R^2 \sin^2 2\varphi_{13}
  + 2 R \sin 2\theta_{23} \sin 2\varphi_{13} \cos\eta \right) \,,
  \\
  \mathcal{R}
  &= \sqrt{1 + R^2 + 2 R \left( \cos 2\theta_{23} \cos 2\varphi_{13}
    + \sin 2\theta_{23} \sin 2\varphi_{13} \cos\eta \right)} \,,
  \\
  R &= \frac{\sqrt{2} G_F N_e(r) \Eps L}{2} \cdot
  \frac{4 E_\nu}{\Dmq_{31}} \,.
\end{align}
So in this case, for large $\Eps$ the unique oscillation wavelength
becomes energy independent in contradiction with the data, and an
upper bound on the magnitude of the non-standard matter effects can be
placed. Also, since in this case we are fixing $\varphi_{12} = \pi/2$
the standard two-neutrino $\nu_\mu \rightarrow \nu_\tau$ oscillation
scenario is recovered in the limit $\Eps = 0$.

\FIGURE[t]{
  \includegraphics[width=0.9\textwidth]{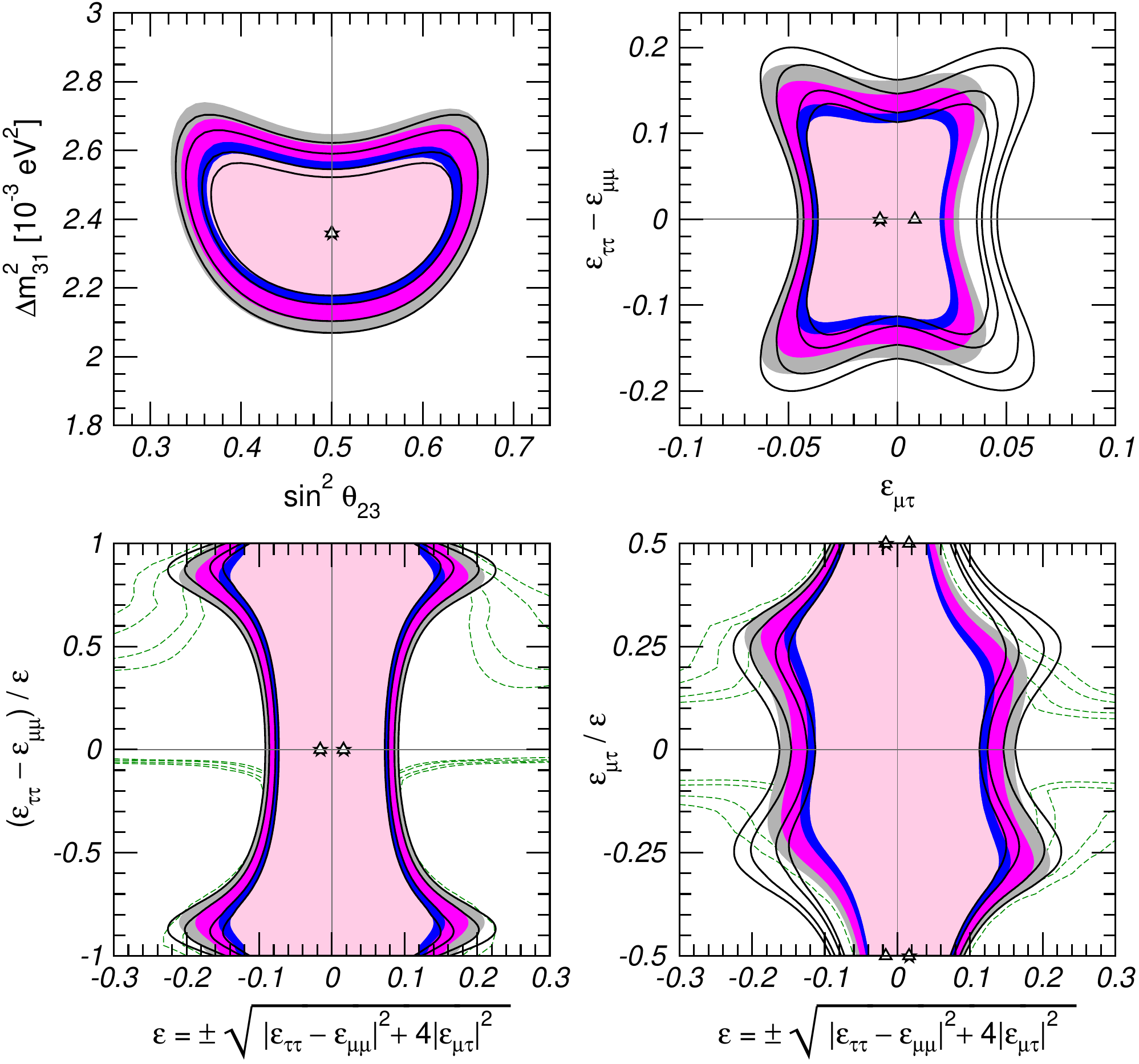}
  \caption{Allowed parameter regions for the analysis of atmospheric
    and LBL data in presence of $\nu_\mu \to \nu_\tau$ oscillations
    and non-standard matter effects. Each panel shows a
    two-dimensional projection of the allowed five-dimensional region
    after marginalization with respect to the three undisplayed
    parameters.  The different contours correspond to the
    two-dimensional allowed regions at 90\%, 95\%, 99\% and $3\sigma$
    CL (2~dof). The filled areas in the upper left panel show the
    projected two-dimensional allowed regions on the ($\Dmq_{31}$,
    $\sin^2\theta_{23}$) plane. The best fit point is marked with a
    star. For the sake of comparison we also show the lines
    corresponding to the contours in the absence of new physics and
    mark with a triangle the position of the best fit point.  The
    regions in other panels show different projections of the allowed
    values of the parameters characterizing the strength and mixing of
    the new matter effects. The full regions correspond to the case
    $\eta \in \lbrace 0,\, \pi \rbrace$ with best fit point marked by
    a star, while the solid lines correspond to arbitrary values of
    the phase $\eta$ with best fit point marked by a triangle.  In the
    two lower panels we show for comparison the corresponding regions
    obtained in the general analysis in Fig.~\ref{fig:range-eps}
    (dashed  contours).
  \label{fig:mutau}
}}

In Fig.~\ref{fig:mutau} we show the two-dimensional projections of the
allowed parameter region for the analysis of atmospheric and LBL
data. The filled areas in the upper left panel show the allowed
regions on the ($\Dmq_{31}$, $\sin^2\theta_{23}$) plane.  For the sake
of comparison we also show the lines corresponding to the contours in
the absence of new physics.  We see that in this constrained scenario
the determination of the oscillation parameters is very much
unaffected by the presence of the non-standard matter effects.  The
other panels of Fig.~\ref{fig:mutau} show different projections of the
allowed values of the parameters characterizing the strength and
mixing of the non-standard matter effects. The solid  contours
correspond to arbitrary values of the phase $\eta$, while the colored
regions correspond to the case $\eta \in \lbrace 0,\, \pi \rbrace$.
Translated in terms of NSI the corresponding 90\% ($3\sigma$) bounds
read:
\begin{equation}
  \label{eq:NSIreallim}
  (-0.055) \; {-0.035} \leq \varepsilon_{\mu\tau} \leq {+0.018} \; (+0.035) \,,
  \qquad
  |\varepsilon_{\tau\tau} - \varepsilon_{\mu\mu}| \leq 0.097 \; (0.16)
\end{equation}
for real NSI, and
\begin{equation}
  \label{eq:NSIcplxlim}
  |\varepsilon_{\mu\tau}| \leq 0.035 \; (0.055) \,,
  \qquad
  |\varepsilon_{\tau\tau}- \varepsilon_{\mu\mu}| \leq 0.11 \; (0.18)
\end{equation}
for the general case of complex $\Eps_{\mu\tau}$.
For the sake of comparison we show in the two lower panels of
Fig.~\ref{fig:mutau} the corresponding regions obtained in the general
analysis in Fig.~\ref{fig:range-eps} as dashed  contours.  We see
how only in the restricted analysis a bound on the $\Eps$ parameter is
set. However the corresponding flavor projections are not more
severely constrained in this restricted analysis when compared to the
general one presented in the previous section.

To finish this section we turn to the possible role of non-vanishing
$\Eps_{\mu\tau}$ and $\Eps_{\tau\tau} - \Eps_{\mu\mu}$ to resolve the
tension between the oscillation results for neutrinos and
antineutrinos in the MINOS experiment~\cite{Adamson:2011ig,
  Adamson:2011fa}.  As shown in the previous section, even in the most
general case we found that the inclusion of the matter potential leads
to a decrease of the $\Delta\chi^2$ of at most $\sim 1.5$ units with
respect to pure oscillations. To further quantify the effect of NSI in
the $\mu\tau$ sector in the MINOS data we show in Fig.~\ref{fig:minos}
(right panel) the $\Delta\chi^2$ of the analysis of MINOS neutrino and
antineutrino results in the framework of $\nu_\mu \rightarrow
\nu_\tau$ oscillations in the presence of non-vanishing NSI in the
$\mu\tau$ sector.  For the sake of comparison with the literature we
show the results for two analysis, one in which all the data of all
energy bins of the MINOS neutrino and antineutrino spectrum are
included and one where only the bins with $1 < E_\nu < 5$~GeV and $1 <
E_{\bar\nu} < 8$~GeV are included as in Ref.~\cite{Kopp:2010qt}.  We
also show in the left panel of Fig.~\ref{fig:minos} the corresponding
oscillation regions for the analysis of MINOS in the framework of pure
$\Dmq_{31}$ $\nu_\mu \rightarrow \nu_\tau$ oscillations for the two
analysis.  As seen in the figure, in the framework of $\Dmq_{31}$
oscillations only both analysis yield equivalent oscillation
regions. This is so because for pure $\Dmq_{31}$ oscillations the data
of the higher energy bins has little relevance in the determination of
the oscillation parameters since at those energies the
$\Dmq_{31}$-driven oscillation wavelength is much longer than the
characteristic $L$ in MINOS so the corresponding survival probability
is 1.
However, when NSI are included, the $\nu_\mu$ and $\bar\nu_\mu$
survival probabilities acquire an energy independent piece as seen in
Eq.~\eqref{eq:prob}.  As a consequence, in the presence of NSI the
inclusion of the high energy bins is relevant to the conclusions of
the analysis.  This is quantified in the right panel of
Fig.~\ref{fig:minos} where we show the $\Delta\chi^2_\text{MINOS}$ for
the five-parameter analysis (in terms of oscillations plus
non-standard matter effects in the $\mu\tau$ sector) as a function of
the strength parameter for the non-standard matter potential after
marginalizing with respect to the other four parameters ($\Dmq_{31}$,
$\theta_{23}$, $\varphi_{13}$, $\eta$). Here
$\Delta\chi^2_\text{MINOS}$ is defined with respect to the pure
oscillation scenario. From this figure we read than when only the low
energy bins are included, NSI seem to be able to improve the agreement
between $\nu_\mu$ and $\bar\nu_\mu$ disappearance data by
$\Delta\chi^2_\text{MINOS} \simeq -7$, in good agreement with the
results in Ref.~\cite{Kopp:2010qt}.  However the inclusion of the
effect of these same NSI in the high energy bins spoils this
improvement and leads only to $\Delta\chi^2_\text{MINOS} \simeq -5$.
Furthermore the required strength for the NSI parameters is in
conflict with the atmospheric results, as shown in the previous
section and correctly noted in~\cite{Kopp:2010qt}.

\FIGURE[t]{
  \includegraphics[width=0.9\textwidth]{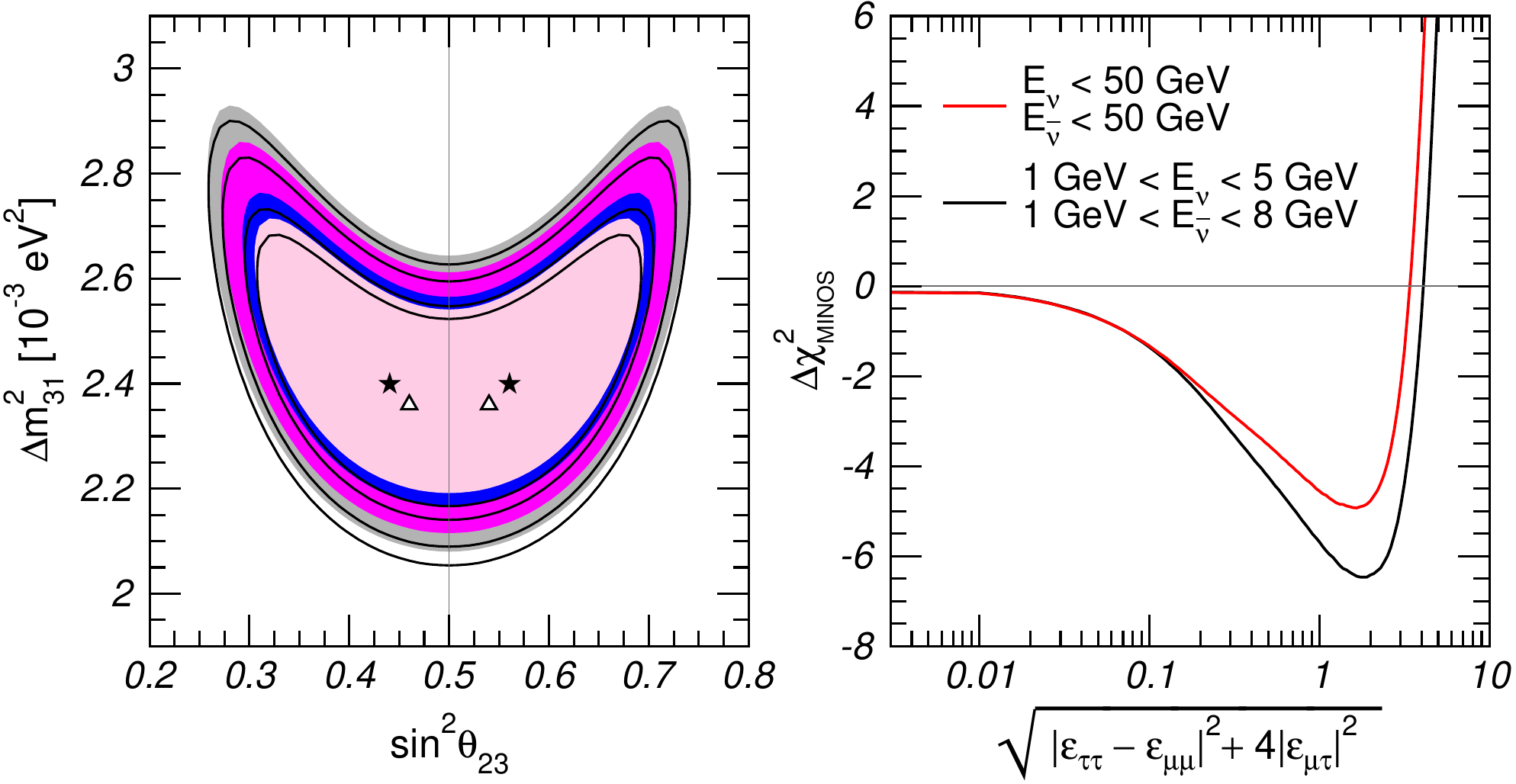}
  \caption{Results of the analysis of MINOS $\nu_\mu$ and
    $\bar\nu_\mu$ disappearance data for two sets of analysis
    differing in the inclusion of the high energy bins (see text for
    details). The left panel show the allowed regions in the framework
    of pure $\Dmq_{31}$-driven oscillations.  The full regions are the
    two-dimensional allowed regions at 90\%, 95\%, 99\% and $3\sigma$
    CL (2~dof) for the analysis including all the bins and the
    corresponding best fit point is marked with a star. The unfilled
    regions correspond to the analysis where only the $1 < E_\nu <
    5$~GeV and $1 < E_{\bar\nu} < 8$~GeV bins are considered (best fit
    point marked with a triangle). The right panel show the
    $\Delta\chi^2_\text{MINOS}$ as a function of the NSI strength for
    the five-parameter analysis (in terms of oscillations plus
    non-standard matter effects in the $\mu\tau$ sector) after
    marginalizing with respect to to the four undisplayed parameters.
    The lower (upper) curve corresponds to the analysis without (with)
    inclusion of the high energy bins.}
  \label{fig:minos}
}

\section{Summary}
\label{sec:conclu}

In this work we have addressed the question: what do atmospheric and
LBL neutrino data teach us about the presence and flavor structure of
matter effects in propagation?

First, to set out to answer this question we have introduced in
Sec.~\ref{sec:forma} a phenomenological parametrization of the matter
potential allowing for departures of the SM one by rescaling of its
strength, rotation from the $ee$ sector, and rephasing with respect to
vacuum.  In this scenario after setting $\Dmq_{21} = 0$ in the
analysis of atmospheric, LBL and CHOOZ data, the relevant flavor
transition probabilities depend on eight parameters: the three
oscillation parameters ($\Dmq_{31}$, $\theta_{23}$ and $\theta_{13}$),
three matter potential parameters ($\Eps$ which represents a rescaling
of the matter potential strength, $\varphi_{12}$ which allows for
projection of the potential into the $\nu_\mu$ flavor, and
$\varphi_{13}$ which allows for its projection into the $\nu_\tau$
flavor) and two relative phases.  The relation to the matter potential
induced by NSI is given in Eq.~\eqref{eq:epsis}.

The answer to the question in the most general case is given in
Sec.~\ref{sec:6par}. We find that in general the strength of the
matter potential $\Eps$ cannot be determined by the analysis.  This is
the generalization of the results in Ref.~\cite{Friedland:2004ah,
  Friedland:2005vy} and it is due to the fact that for $\sqrt{2} G_F
N_e(r) \Eps \gg \Dmq_{31} / (2 E_\nu)$, $\nu_\mu\leftrightarrow
\nu_\tau$ transitions oscillate dominantly with the phase
$\tilde{\Delta}_\text{vac}$ in Eq.~\eqref{eq:Dvac} and have the same
dependence on the neutrino energy and distance as vacuum
oscillations. This allows for good description of atmospheric and LBL
results even for very large values of $\Eps$.

However important information is obtained from the data analysis on
the flavor structure of the matter potential. This is so because for
large $\Eps$ the $\nu_\mu\leftrightarrow \nu_e$ (Eq.~\eqref{eq:Pem})
and $\nu_e\leftrightarrow \nu_\tau$ (Eq.~\eqref{eq:Pet}) transitions
proceed dominantly via oscillations with the energy independent phase
$\tilde{\Delta}_\text{mat}$ in Eqs.~\eqref{eq:Dmat} and amplitudes
controlled by the projection angles $\varphi_{12}$ and
$\varphi_{13}$. This means that even for $\theta_{13}=0$ atmospheric
$\nu_e$'s can disappear into either $\nu_\mu$ or $\nu_\tau$ and they
do so independently of their energy. Similarly, some atmospheric and
LBL $\nu_\mu$'s will oscillate into electron neutrinos independently
of their energy. This is in clear conflict with the atmospheric and
LBL data, and as a consequence potentials with $|\Eps| \gtrsim
\mathcal{O}(0.2)$ are only allowed as long as their flavor projections
out of the $ee$ entry are severely constrained. As can be seen in
Fig.~\ref{fig:proj-eps}, the constraints are stronger on rotations
over the $\mu$ flavor, controlled by $\varphi_{12}$.

We also find that the present determination of the oscillation
parameters is rather robust even in the presence of this general form
of the matter potential as illustrated in Figs.~\ref{fig:proj-osc}
and~\ref{fig:range-osc}.  For example the $3\sigma$ ranges of the
oscillation parameters read:
\begin{equation}
  \label{eq:oscrang}
  \begin{array}{cc}
    \text{Oscillations}
    & \text{Oscillations} + \text{Generalized Matter Potential}
    \\[1mm]
    2.11 \times 10^{-3} \leq |\Dmq_{31}| \leq 2.69 \times 10^{-3}
    & 2.10 \times 10^{-3} \leq |\Dmq_{31}| \leq 3.06 \times 10^{-3}
    \\[1mm]
    0.34 \leq \sin^2\theta_{23}\leq 0.68
    & 0.26 \leq \sin^2\theta_{23}\leq 0.68
    \\[1mm]
    \hspace{14mm} \sin^2\theta_{13}\leq 0.051
    & \hspace{14mm} \sin^2\theta_{13}\leq 0.060
  \end{array}
\end{equation}

Concerning the quality of the fit, we find that even though no
constraint on the magnitude of the matter effects can be placed by the
data, the data does not show any favoring for a non-standard matter
potential. There is no substantial improvement on the statistical
quality of the fit due to the inclusion of the extra matter parameters
with $\Delta\chi^2$ reduced by a maximum of $1.5$ units in the global
analysis. This is so even after the inclusion of the MINOS
$\bar\nu_\mu$ disappearance results. Thus we conclude that within the
context of a global analysis the inclusion of the NSI does not
significantly alleviate the tension between the oscillation results
for neutrinos and antineutrinos arising from the MINOS $\bar\nu_\mu$
disappearance results.  This is partly due to the constraints from
atmospheric neutrino data and partly to the fact that the non-standard
matter potential parameters required to better fit the MINOS energy
spectrum of $\nu_\mu$ and $\bar\nu_\mu$ events with $E_\nu\leq 5$ GeV
and $E_{\bar\nu}\leq 8$ GeV with the same oscillation parameters
worsens the description of the spectrum of $E_\nu> 5$ GeV events as
shown in the right panel of Fig.~\ref{fig:minos}.

Finally we have revisited the particular scenario in which matter
effects in the $e\mu$ and $e\tau$ sectors are very suppressed
(Sec.~\ref{sec:4par}).  In this case a bound on the strength of the
NSI in the $\mu\tau$ sector can be set. The corresponding updated
bounds shown in Eqs.~\eqref{eq:NSIreallim} and~\eqref{eq:NSIcplxlim}
are 5-20\% stronger than our last published bounds in
Ref.~\cite{GonzalezGarcia:2007ib}.

\section*{Acknowledgments}

We thank O.~Peres for comments.  This work is supported by Spanish
MICINN grants 2007-66665-C02-01, FPA-2009-08958, FPA-2009-09017 and
consolider-ingenio 2010 grant CSD-2008-0037, by CUR Generalitat de
Catalunya grant 2009SGR502, by Comunidad Autonoma de Madrid through
the HEPHACOS project S2009/ESP-1473, by USA-NSF grant PHY-0969739 and
by EU grant EURONU.

\appendix

\section{Parametrization of the matter potential}
\label{sec:app1}

We give here some details on the simplifications associated with the
condition that both the vacuum and the matter part of the Hamiltonian
have each two degenerate eigenvalues. In the most general case the
matter potential can be parametrized as:
\begin{equation}
  \label{eq:hmatgen}
  H_\text{mat} = Q_\text{rel} U_\text{mat} D_\text{mat}
  U_\text{mat}^\dagger Q_\text{rel}^\dagger
  \text{~~with~~}
  \left\lbrace
  \begin{aligned}
    Q_\text{rel} &= \diag\left(
    e^{i\alpha_1}, e^{i\alpha_2}, e^{-i\alpha_1 -i\alpha_2} \right) ,
    \\
    U_\text{mat} &= R_{12}(\varphi_{12})
    \tilde{R}_{13}(\varphi_{13}, \delta_\text{NS})
    R_{23}(\varphi_{23}) \,,
    \\
    D_\text{mat} &= \sqrt{2} G_F N_e(r) \diag(\Eps, \Eps', 0) \,.
  \end{aligned}\right.
\end{equation}
Setting $\Eps' = 0$ implies that the $\varphi_{23}$ angle and the
$\delta_\text{NS}$ phase become unphysical. In the limit $\Dmq_{21} =
0$ the general Hamiltonian reduces to:
\begin{equation}
  H = R_{23}^\theta R_{13}^\theta D_\text{vac}
  R_{13}^{\theta\dagger} R_{23}^{\theta\dagger}
  + Q_\text{rel} R_{12}^\varphi R_{13}^\varphi D_\text{mat}
  R_{13}^{\varphi\dagger} R_{12}^{\varphi\dagger} Q_\text{rel}^\dagger
\end{equation}
where $R_{ij}^\theta \equiv R_{ij}(\theta_{ij})$ and $R_{ij}^\varphi \equiv
R_{ij}(\varphi_{ij})$. We can write:
\begin{equation}
  H = R_{23}^\theta R_{13}^\theta \left[ D_\text{vac} + \left(
    R_{13}^{\theta\dagger} R_{23}^{\theta\dagger}
    Q_\text{rel} R_{12}^\varphi R_{13}^\varphi \right)
    D_\text{mat} \left( R_{13}^{\varphi\dagger} R_{12}^{\varphi\dagger}
    Q_\text{rel}^\dagger R_{23}^\theta R_{13}^\theta
    \right) \right] R_{13}^{\theta\dagger} R_{23}^{\theta\dagger}
\end{equation}
where we simply collected the two real rotations $R_{23}^\theta$ and
$R_{13}^\theta$. Changing the parametrization:
\begin{equation}
  R_{13}^{\theta\dagger} R_{23}^{\theta\dagger}
  Q_\text{rel} R_{12}^\varphi R_{13}^\varphi
  \to Q_\text{eff} R_{12}^\psi R_{13}^\psi
  \tilde{R}_{23}^{\psi,\delta} Q_\text{maj}
\end{equation}
where $Q_\text{eff} = \diag\left( e^{i\beta_1}, e^{i\beta_2},
e^{-i\beta_1 -i\beta_2} \right)$ are two new relative phases,
$R_{12}^\psi$ and $R_{13}^\psi$ are \emph{real} rotations by angles
$\psi_{12}$ and $\psi_{13}$, $\tilde{R}_{23}^{\psi,\delta}$ is a
\emph{complex} rotation by angle $\psi_{23}$ and phase $\delta_{23}$,
and $Q_\text{maj}$ contains two Majorana phases. The matrix on the
\emph{l.h.s.}\ is a $\det = 1$ unitary matrix, so it can be decomposed
as in the \emph{r.h.s.}. The new angles ($\psi_{12}$, $\psi_{13}$,
$\psi_{23}$) as well as the phases ($\delta_{23}$, $\beta_1$,
$\beta_2$) and the Majorana phases in $Q_\text{maj}$ are some
complicated functions of the original angles ($\theta_{13}$,
$\theta_{23}$, $\varphi_{12}$, $\varphi_{13}$) and of the phases
($\alpha_1$, $\alpha_2$), but they \emph{do not depend} on
$\Dmq_{31}$, $E_\nu$ (neutrino energy), $\Eps$, or $N_e(r)$. They are
universal and independent of the details of the trajectory.

Now, obviously the Majorana phases $Q_\text{maj}$ disappear, as well
as $\delta_{23}$ and $\psi_{23}$ (since they rotate the $(0, 0)$ block
in $D_\text{mat}$). Hence we are left with:
\begin{equation}
  H = R_{23}^\theta R_{13}^\theta \left[ D_\text{vac} +
    Q_\text{eff} R_{12}^\psi R_{13}^\psi D_\text{mat} R_{13}^{\psi\dagger}
    R_{12}^{\psi\dagger} Q_\text{eff}^\dagger \right]
  R_{13}^{\theta\dagger} R_{23}^{\theta\dagger}
\end{equation}
Now we note that:
\begin{equation}
  D_\text{vac} = Q_\text{eff} R_{12}^\psi D_\text{vac}
  R_{12}^{\psi\dagger} Q_\text{eff}^\dagger
\end{equation}
since $R_{12}^\psi$ rotates the $(0, 0)$ block in $D_\text{vac}$ and
$Q_\text{eff}$ is diagonal. Therefore:
\begin{equation}
  H = U_\text{eff} \left[D_\text{vac} +
    R_{13}^\psi D_\text{mat} R_{13}^{\psi\dagger} \right] U_\text{eff}^\dagger
  \quad\text{where}\quad
  U_\text{eff} \equiv R_{23}^\theta R_{13}^\theta Q_\text{eff} R_{12}^\psi
\end{equation}
This lead to a scattering matrix:
\begin{equation}
  S = U_\text{eff} S_\text{eff} U_\text{eff}^\dagger
  \quad\text{with}\quad
  S_\text{eff} = \text{evolution-of} \left(
  D_\text{vac} + R_{13}^\psi D_\text{mat} R_{13}^{\psi\dagger} \right)
\end{equation}
It is immediate to see that the effective Hamiltonian $H_\text{eff} =
D_\text{vac} + R_{13}^\psi D_\text{mat} R_{13}^{\psi\dagger}$ depends
only on 3 parameters ($\Dmq_{31}$, $\Eps$, $\psi_{13}$) and has the
form (up to an overall multiple of the identity):
\begin{equation}
  H_\text{eff} =
  \begin{pmatrix}
    H_D & 0 & H_N \\
    0 & H_P & 0 \\
    H_N & 0 & -H_{D}
  \end{pmatrix}
  \qquad\Longrightarrow\qquad
  S_\text{eff} =
  \begin{pmatrix}
    S_D & 0 & -S_N^* \\
    0 & S_P & 0 \\
    S_N & 0 & S_D^*
  \end{pmatrix}
\end{equation}
hence it factorizes into a $2\times 2$ block and a $1\times 1$
block. The remaining 5 parameters $\theta_{13}$, $\theta_{23}$,
$\beta_1$, $\beta_2$, $\psi_{12}$ are non-dynamical and can be
reinserted later to form the $\chi^2$. Concretely, the neutrino
probabilities can be written as the sum of 18 different terms:
\begin{equation}
  P_{\alpha\beta} = \sum_{n=1}^{18} C^{\alpha\beta}_n P^\text{eff}_n
\end{equation}
where $C^{\alpha\beta}_n = C^{\alpha\beta}_n(\theta_{13}, \theta_{23},
\beta_1, \beta_2, \psi_{12})$ and $P^\text{eff}_n =
P^\text{eff}_n(\Dmq_{31}, \Eps, \psi_{13})$ are real numbers related
to $U_\text{eff}$ and $S_\text{eff}$ by the following formulas:
\begin{equation}
  \label{eq:coeff}
  \begin{aligned}
    C^{\alpha\beta}_{1}
    &= |W^{\alpha\beta}_{22}|^2 + |W^{\alpha\beta}_{31}|^2
    + |W^{\alpha\beta}_{13}|^2 \,,
    & P^\text{eff}_{1}
    &= 1 \,,
    \\
    C^{\alpha\beta}_{2}
    &= |W^{\alpha\beta}_{11}|^2 + |W^{\alpha\beta}_{33}|^2
    - |W^{\alpha\beta}_{13}|^2 - |W^{\alpha\beta}_{31}|^2 \,,
    & P^\text{eff}_{2}
    &= |S_D|^2 \,,
    \\
    C^{\alpha\beta}_{3} + i C^{\alpha\beta}_{4}
    &= \hphantom{-} {W^{\alpha\beta}_{11}}^* W^{\alpha\beta}_{31}
    - {W^{\alpha\beta}_{13}}^* W^{\alpha\beta}_{33} \,,
    & P^\text{eff}_{3} + i P^\text{eff}_{4}
    &= 2 \, S_D \, S_N^* \,,
    \\
    C^{\alpha\beta}_{5} + i C^{\alpha\beta}_{6}
    &= \hphantom{-} {W^{\alpha\beta}_{11}}^* W^{\alpha\beta}_{22} \,,
    & P^\text{eff}_{5} + i P^\text{eff}_{6}
    &= 2 \, S_D \, S_P^* \,,
    \\
    C^{\alpha\beta}_{7} + i C^{\alpha\beta}_{8}
    &= \hphantom{-} {W^{\alpha\beta}_{31}}^* W^{\alpha\beta}_{22} \,,
    & P^\text{eff}_{7} + i P^\text{eff}_{8}
    &= 2 \, S_N \, S_P^* \,,
    \\
    C^{\alpha\beta}_{9} + i C^{\alpha\beta}_{10}
    &= \hphantom{-} {W^{\alpha\beta}_{11}}^* W^{\alpha\beta}_{33} \,,
    & P^\text{eff}_{9} + i P^\text{eff}_{10}
    &= 2 \, S_D \, S_D \,,
    \\
    C^{\alpha\beta}_{11} + i C^{\alpha\beta}_{12}
    &= -{W^{\alpha\beta}_{31}}^* W^{\alpha\beta}_{13} \,,
    & P^\text{eff}_{11} + i P^\text{eff}_{12}
    &= 2 \, S_N \, S_N \,,
    \\
    C^{\alpha\beta}_{13} + i C^{\alpha\beta}_{14}
    &= \hphantom{-} {W^{\alpha\beta}_{31}}^* W^{\alpha\beta}_{33}
    - {W^{\alpha\beta}_{11}}^* W^{\alpha\beta}_{13} \,,
    & P^\text{eff}_{13} + i P^\text{eff}_{14}
    &= 2 \, S_D \, S_N \,,
    \\
    C^{\alpha\beta}_{15} + i C^{\alpha\beta}_{16}
    &= \hphantom{-} {W^{\alpha\beta}_{22}}^* W^{\alpha\beta}_{33} \,,
    & P^\text{eff}_{15} + i P^\text{eff}_{16}
    &= 2 \, S_D \, S_P \,,
    \\
    C^{\alpha\beta}_{17} + i C^{\alpha\beta}_{18}
    &= -{W^{\alpha\beta}_{22}}^* W^{\alpha\beta}_{13} \,,
    & P^\text{eff}_{17} + i P^\text{eff}_{18}
    &= 2 \, S_N \, S_P \,,
  \end{aligned}
\end{equation}
with $W^{\alpha\beta}_{ij} \equiv U^\text{eff}_{\alpha i} \,
{U^\text{eff}_{\beta j}}^*$. For antineutrinos one has to flip the
sign of the matter term in $H_\text{eff}$ and replace $U_\text{eff}
\to U_\text{eff}^*$ (or, equivalently, $S_\text{eff} \to
S_\text{eff}^*$) in Eq.~\eqref{eq:coeff}.

\section{Symmetries and parameter ranges}
\label{sec:app2}

In order to describe the physical ranges for the parameters in the
problem we study which kind of transformations leave the probabilities
unaffected.  Let's consider a problem described by an Hamiltonian
$H$. The transformation $H \to -H^*$ implies $\exp(-iHL) \to \exp(iH^*
L) = \exp[(-iHL)^*] = [\exp(-iHL)]^*$, hence it leaves the
probabilities unchanged. Such transformation is realized by:
\begin{enumerate}
\item[(0)] $\Dmq_{31} \to -\Dmq_{31} ~\wedge~ \Eps \to -\Eps
  ~\wedge~ \alpha_1 \to -\alpha_1 ~\wedge~ \alpha_2 \to -\alpha_2$;
\end{enumerate}
This symmetry implies that only the \emph{relative} sign of
$\Dmq_{31}$ and $\Eps$ matters, so that we can either assume $\Eps \ge
0$ and study both signs of $\Dmq_{31}$, or assume $\Dmq_{31} \ge 0$
and study both sign of $\Eps$. Furthermore, any rephasing $H \to Q H
Q^*$ where $Q = \diag\left( e^{ia}, e^{ib}, e^{ic} \right)$ leads to a
rephasing of the scattering matrix $\exp(-iHL) \to Q \exp(-iHL) Q^*$,
which leaves the probabilities unaffected. Our Hamiltonian can be
written as:
\begin{equation}
  H = U_\text{vac} D_\text{vac} U_\text{vac}^\dagger
  + Q_\text{rel} U_\text{mat} D_\text{mat}
  U_\text{mat}^\dagger Q_\text{rel}^\dagger
\end{equation}
where $U_\text{vac} = R_{23}(\theta_{23}) R_{13}(\theta_{13})$ and
$U_\text{mat} = R_{12}(\varphi_{12}) R_{13}(\varphi_{13})$. A
transformation $H \to Q H Q^*$ is achieved if:
\begin{equation}
  U_\text{vac} \to Q U_\text{vac} Q'
  \quad\text{and}\quad
  Q_\text{rel} U_\text{mat} \to Q Q_\text{rel} U_\text{mat} Q''
\end{equation}
where $Q$, $Q'$, $Q''$ are all pure phase diagonal matrices. Hence,
any transformation of the parameters which is equivalent to a specific
choice of $Q$, $Q'$ and $Q''$ will leave the probabilities
unaffected. With this, we can generate a list of symmetries:
\begin{enumerate}
\item[(1)] $\theta_{13} \to \pi + \theta_{13}$. This is equivalent to
  $Q' = \diag(-1, +1, -1)$;

\item[(2)] $\theta_{13} \to -\theta_{13} ~\wedge~ \theta_{23} \to \pi
  + \theta_{23}$. This is equivalent to $Q' = \diag(+1, -1, -1)$;

\item[(3)] $\varphi_{13} \to \pi + \varphi_{13}$. This is equivalent
  to $Q'' = \diag(-1, +1, -1)$;

\item[(4)] $\varphi_{13} \to -\varphi_{13} ~\wedge~ \varphi_{12} \to
  \pi + \varphi_{12}$. This is equivalent to $Q'' = \diag(-1, -1,
  +1)$;

\item[(5)] $\theta_{23} \to -\theta_{23} ~\wedge~ \varphi_{12} \to
  -\varphi_{12}$. This is equivalent to $Q = Q' = Q'' = \diag(+1, -1,
  +1)$;

\item[(6)] $\theta_{23} \to \pi - \theta_{23} ~\wedge~ \varphi_{12}
  \to \pi + \varphi_{12}$. This is equivalent to $Q = \diag(+1, +1,
  -1)$, $Q' = \diag(+1, -1, +1)$, $Q'' = -I$;

\item[(7)] $\varphi_{12} \to -\varphi_{12} ~\wedge~ \alpha_1 \to
  \alpha_1 + \pi/3 ~\wedge~ \alpha_2 \to \alpha_2 - 2\pi/3$. This is
  equivalent to $Q'' = e^{i\pi/3} \diag(+1, -1, +1)$.

\item[(8)] $\varphi_{12} \to \pi + \varphi_{12} ~\wedge~ \alpha_1 \to
  \alpha_1 + \pi/3 ~\wedge~ \alpha_2 \to \alpha_2 + \pi/3$. This is
  equivalent to $Q'' = e^{-2i\pi/3} I$.
\end{enumerate}
With these symmetries we can reduce the range of the parameters
($\theta_{23}$, $\theta_{13}$, $\varphi_{12}$, $\varphi_{13}$) from
the most general $[0 \to 2\pi]$ to something less:
\begin{itemize}
\item using (1) we can reduce $\theta_{13}$ to the range $[-\pi/2 \to
  +\pi/2]$;

\item using (2) we can reduce $\theta_{13}$ to the range $[0 \to
  \pi/2]$;

\item using (5) we can reduce $\theta_{23}$ to the range $[0 \to
  \pi]$;

\item using (6) we can reduce $\theta_{23}$ to the range $[0 \to
  \pi/2]$;

\item using (3) we can reduce $\varphi_{13}$ to the range $[-\pi/2 \to
  \pi/2]$.
\end{itemize}
For the case of \textbf{real NSI}:
\begin{itemize}
\item using (4) we can reduce $\varphi_{12}$ to the range $[-\pi/2 \to
  \pi/2]$;
\end{itemize}
so that for $\alpha_1 = \alpha_2 = 0$ we can reduce the parameter
range for the angles to $0 < \theta_{ij}< \pi/2$ and $-\pi/2 <
\varphi_{ij} < \pi/2$.
For the case of \textbf{complex NSI}:
\begin{itemize}
\item using (4) we can reduce $\varphi_{13}$ to the range $[0 \to
  \pi/2]$;

\item using (8) we can reduce $\varphi_{12}$ to the range $[-\pi/2 \to
  \pi/2]$;

\item using (7) we can reduce $\varphi_{12}$ to the range $[0 \to
  \pi/2]$;
\end{itemize}
so that in the general case of unconstrained $\alpha_i$ we can reduce
the parameter range for the angles to $0 < \theta_{ij} < \pi/2$ and $0
< \varphi_{ij} < \pi/2$.

\bibliographystyle{JHEP}
\bibliography{references}

\providecommand{\href}[2]{#2}\begingroup\raggedright\begin{thebibliography}{10}

\bibitem{Pontecorvo:1967fh}
B.~Pontecorvo, {\it {Neutrino experiments and the question of leptonic-charge
  conservation}},  {\em Sov. Phys. JETP} {\bf 26} (1968) 984--988.

\bibitem{Gribov:1968kq}
V.~N. Gribov and B.~Pontecorvo, {\it {Neutrino astronomy and lepton charge}},
  {\em Phys. Lett.} {\bf B28} (1969) 493.

\bibitem{GonzalezGarcia:2007ib}
M.~C. Gonzalez-Garcia and M.~Maltoni, {\it {Phenomenology with Massive
  Neutrinos}},  {\em Phys. Rept.} {\bf 460} (2008) 1--129,
  [\href{http://arxiv.org/abs/0704.1800}{{\tt arXiv:0704.1800}}].

\bibitem{Maki:1962mu}
Z.~Maki, M.~Nakagawa, and S.~Sakata, {\it {Remarks on the unified model of
  elementary particles}},  {\em Prog. Theor. Phys.} {\bf 28} (1962) 870--880.

\bibitem{Kobayashi:1973fv}
M.~Kobayashi and T.~Maskawa, {\it {CP Violation in the Renormalizable Theory of
  Weak Interaction}},  {\em Prog. Theor. Phys.} {\bf 49} (1973) 652--657.

\bibitem{Bilenky:1980cx}
S.~M. Bilenky, J.~Hosek, and S.~T. Petcov, {\it {On Oscillations of Neutrinos
  with Dirac and Majorana Masses}},  {\em Phys. Lett.} {\bf B94} (1980) 495.

\bibitem{Langacker:1986jv}
P.~Langacker, S.~T. Petcov, G.~Steigman, and S.~Toshev, {\it {On the
  Mikheev-Smirnov-Wolfenstein (MSW) Mechanism of Amplification of Neutrino
  Oscillations in Matter}},  {\em Nucl. Phys.} {\bf B282} (1987) 589.

\bibitem{Schwetz:2011qt}
T.~Schwetz, M.~Tortola, and J.~Valle, {\it {Global neutrino data and recent
  reactor fluxes: status of three-flavour oscillation parameters}},
  \href{http://arxiv.org/abs/1103.0734}{{\tt arXiv:1103.0734}}.

\bibitem{GonzalezGarcia:2010er}
M.~C. Gonzalez-Garcia, M.~Maltoni, and J.~Salvado, {\it {Updated global fit to
  three neutrino mixing: status of the hints of theta13 > 0}},  {\em JHEP} {\bf
  04} (2010) 056, [\href{http://arxiv.org/abs/1001.4524}{{\tt
  arXiv:1001.4524}}].

\bibitem{Fogli:2009zza}
G.~L. Fogli, E.~Lisi, A.~Marrone, A.~Palazzo, and A.~M. Rotunno, {\it {Neutrino
  masses and mixing: 2008 status}},  {\em Nucl. Phys. Proc. Suppl.} {\bf 188}
  (2009) 27--30.

\bibitem{Maltoni:2008ka}
M.~Maltoni and T.~Schwetz, {\it {Three-flavour neutrino oscillation update and
  comments on possible hints for a non-zero $\theta_{13}$}},  {\em PoS} {\bf
  IDM2008} (2008) 072, [\href{http://arxiv.org/abs/0812.3161}{{\tt
  arXiv:0812.3161}}].

\bibitem{Wolfenstein:1977ue}
L.~Wolfenstein, {\it {Neutrino Oscillations in Matter}},  {\em Phys.Rev.} {\bf
  D17} (1978) 2369--2374.

\bibitem{Mikheev:1986gs}
S.~Mikheev and A.~Smirnov, {\it {Resonance Amplification of Oscillations in
  Matter and Spectroscopy of Solar Neutrinos}},  {\em Sov.J.Nucl.Phys.} {\bf
  42} (1985) 913--917.

\bibitem{Fogli:2002hb}
G.~Fogli, E.~Lisi, A.~Palazzo, and A.~Rotunno, {\it {Solar neutrino
  oscillations and indications of matter effects in the sun}},  {\em Phys.Rev.}
  {\bf D67} (2003) 073001, [\href{http://arxiv.org/abs/hep-ph/0211414}{{\tt
  hep-ph/0211414}}].

\bibitem{Roulet:1991sm}
E.~Roulet, {\it {MSW effect with flavor changing neutrino interactions}},  {\em
  Phys.Rev.} {\bf D44} (1991) 935--938.

\bibitem{Guzzo:1991hi}
M.~Guzzo, A.~Masiero, and S.~Petcov, {\it {On the MSW effect with massless
  neutrinos and no mixing in the vacuum}},  {\em Phys.Lett.} {\bf B260} (1991)
  154--160.

\bibitem{Barger:1991ae}
V.~D. Barger, R.~Phillips, and K.~Whisnant, {\it {Solar neutrino solutions with
  matter enhanced flavor changing neutral current scattering}},  {\em
  Phys.Rev.} {\bf D44} (1991) 1629--1643.

\bibitem{Fogli:1993xv}
G.~L. Fogli and E.~Lisi, {\it {Solar Neutrino data, solar model uncertainties
  and solar matter enhanced neutrino oscillations}},  {\em Astropart.Phys.}
  {\bf 2} (1994) 91--100.

\bibitem{Bergmann:1997mr}
S.~Bergmann, {\it {The Solar neutrino problem in the presence of flavor
  changing neutrino interactions}},  {\em Nucl.Phys.} {\bf B515} (1998)
  363--383, [\href{http://arxiv.org/abs/hep-ph/9707398}{{\tt hep-ph/9707398}}].

\bibitem{Bergmann:2000gp}
S.~Bergmann, M.~Guzzo, P.~de~Holanda, P.~Krastev, and H.~Nunokawa, {\it {Status
  of the solution to the solar neutrino problem based on nonstandard neutrino
  interactions}},  {\em Phys.Rev.} {\bf D62} (2000) 073001,
  [\href{http://arxiv.org/abs/hep-ph/0004049}{{\tt hep-ph/0004049}}].

\bibitem{Guzzo:2000kx}
M.~Guzzo, H.~Nunokawa, P.~de~Holanda, and O.~Peres, {\it {On the massless
  'just-so' solution to the solar neutrino problem}},  {\em Phys.Rev.} {\bf
  D64} (2001) 097301, [\href{http://arxiv.org/abs/hep-ph/0012089}{{\tt
  hep-ph/0012089}}].

\bibitem{Friedland:2004pp}
A.~Friedland, C.~Lunardini, and C.~Pena-Garay, {\it {Solar neutrinos as probes
  of neutrino matter interactions}},  {\em Phys.Lett.} {\bf B594} (2004) 347,
  [\href{http://arxiv.org/abs/hep-ph/0402266}{{\tt hep-ph/0402266}}].

\bibitem{Escrihuela:2009up}
F.~Escrihuela, O.~Miranda, M.~Tortola, and J.~Valle, {\it {Constraining
  nonstandard neutrino-quark interactions with solar, reactor and accelerator
  data}},  {\em Phys.Rev.} {\bf D80} (2009) 105009,
  [\href{http://arxiv.org/abs/0907.2630}{{\tt arXiv:0907.2630}}].

\bibitem{Bolanos:2008km}
A.~Bolanos, O.~Miranda, A.~Palazzo, M.~Tortola, and J.~Valle, {\it {Probing
  non-standard neutrino-electron interactions with solar and reactor
  neutrinos}},  {\em Phys.Rev.} {\bf D79} (2009) 113012,
  [\href{http://arxiv.org/abs/0812.4417}{{\tt arXiv:0812.4417}}].

\bibitem{Minakata:2010be}
H.~Minakata and C.~Pena-Garay, {\it {Solar Neutrino Observables Sensitive to
  Matter Effects}},  \href{http://arxiv.org/abs/1009.4869}{{\tt
  arXiv:1009.4869}}.

\bibitem{Palazzo:2011vg}
A.~Palazzo, {\it {Hint of non-standard MSW dynamics in solar neutrino
  conversion}},  \href{http://arxiv.org/abs/1101.3875}{{\tt arXiv:1101.3875}}.

\bibitem{Grossman:1995wx}
Y.~Grossman, {\it {Nonstandard neutrino interactions and neutrino oscillation
  experiments}},  {\em Phys.Lett.} {\bf B359} (1995) 141--147,
  [\href{http://arxiv.org/abs/hep-ph/9507344}{{\tt hep-ph/9507344}}].

\bibitem{GonzalezGarcia:2001mp}
M.~Gonzalez-Garcia, Y.~Grossman, A.~Gusso, and Y.~Nir, {\it {New CP violation
  in neutrino oscillations}},  {\em Phys.Rev.} {\bf D64} (2001) 096006,
  [\href{http://arxiv.org/abs/hep-ph/0105159}{{\tt hep-ph/0105159}}].

\bibitem{Gago:2001xg}
A.~Gago, M.~Guzzo, H.~Nunokawa, W.~Teves, and R.~Zukanovich~Funchal, {\it
  {Probing flavor changing neutrino interactions using neutrino beams from a
  muon storage ring}},  {\em Phys.Rev.} {\bf D64} (2001) 073003,
  [\href{http://arxiv.org/abs/hep-ph/0105196}{{\tt hep-ph/0105196}}].

\bibitem{Fornengo:2001pm}
N.~Fornengo, M.~Maltoni, R.~Tomas, and J.~Valle, {\it {Probing neutrino
  nonstandard interactions with atmospheric neutrino data}},  {\em Phys.Rev.}
  {\bf D65} (2002) 013010, [\href{http://arxiv.org/abs/hep-ph/0108043}{{\tt
  hep-ph/0108043}}].

\bibitem{Huber:2001zw}
P.~Huber and J.~Valle, {\it {Nonstandard interactions: Atmospheric versus
  neutrino factory experiments}},  {\em Phys.Lett.} {\bf B523} (2001) 151--160,
  [\href{http://arxiv.org/abs/hep-ph/0108193}{{\tt hep-ph/0108193}}].

\bibitem{Ota:2001pw}
T.~Ota, J.~Sato, and N.-a. Yamashita, {\it {Oscillation enhanced search for new
  interaction with neutrinos}},  {\em Phys.Rev.} {\bf D65} (2002) 093015,
  [\href{http://arxiv.org/abs/hep-ph/0112329}{{\tt hep-ph/0112329}}].

\bibitem{Huber:2002bi}
P.~Huber, T.~Schwetz, and J.~Valle, {\it {Confusing nonstandard neutrino
  interactions with oscillations at a neutrino factory}},  {\em Phys.Rev.} {\bf
  D66} (2002) 013006, [\href{http://arxiv.org/abs/hep-ph/0202048}{{\tt
  hep-ph/0202048}}].

\bibitem{Campanelli:2002cc}
M.~Campanelli and A.~Romanino, {\it {Effects of new physics in neutrino
  oscillations in matter}},  {\em Phys.Rev.} {\bf D66} (2002) 113001,
  [\href{http://arxiv.org/abs/hep-ph/0207350}{{\tt hep-ph/0207350}}].

\bibitem{Ota:2002na}
T.~Ota and J.~Sato, {\it {Can ICARUS and OPERA give information on a new
  physics?}},  {\em Phys.Lett.} {\bf B545} (2002) 367--372,
  [\href{http://arxiv.org/abs/hep-ph/0202145}{{\tt hep-ph/0202145}}].

\bibitem{GonzalezGarcia:2004wg}
M.~Gonzalez-Garcia and M.~Maltoni, {\it {Atmospheric neutrino oscillations and
  new physics}},  {\em Phys.Rev.} {\bf D70} (2004) 033010,
  [\href{http://arxiv.org/abs/hep-ph/0404085}{{\tt hep-ph/0404085}}].

\bibitem{Friedland:2004ah}
A.~Friedland, C.~Lunardini, and M.~Maltoni, {\it {Atmospheric neutrinos as
  probes of neutrino-matter interactions}},  {\em Phys.Rev.} {\bf D70} (2004)
  111301, [\href{http://arxiv.org/abs/hep-ph/0408264}{{\tt hep-ph/0408264}}].

\bibitem{Friedland:2005vy}
A.~Friedland and C.~Lunardini, {\it {A Test of tau neutrino interactions with
  atmospheric neutrinos and K2K}},  {\em Phys.Rev.} {\bf D72} (2005) 053009,
  [\href{http://arxiv.org/abs/hep-ph/0506143}{{\tt hep-ph/0506143}}].

\bibitem{Blennow:2005qj}
M.~Blennow, T.~Ohlsson, and W.~Winter, {\it {Non-standard Hamiltonian effects
  on neutrino oscillations}},  {\em Eur.Phys.J.} {\bf C49} (2007) 1023--1039,
  [\href{http://arxiv.org/abs/hep-ph/0508175}{{\tt hep-ph/0508175}}].

\bibitem{Kitazawa:2006iq}
N.~Kitazawa, H.~Sugiyama, and O.~Yasuda, {\it {Will MINOS see new physics?}},
  \href{http://arxiv.org/abs/hep-ph/0606013}{{\tt hep-ph/0606013}}.

\bibitem{Friedland:2006pi}
A.~Friedland and C.~Lunardini, {\it {Two modes of searching for new neutrino
  interactions at MINOS}},  {\em Phys.Rev.} {\bf D74} (2006) 033012,
  [\href{http://arxiv.org/abs/hep-ph/0606101}{{\tt hep-ph/0606101}}].

\bibitem{Blennow:2007pu}
M.~Blennow, T.~Ohlsson, and J.~Skrotzki, {\it {Effects of non-standard
  interactions in the MINOS experiment}},  {\em Phys.Lett.} {\bf B660} (2008)
  522--528, [\href{http://arxiv.org/abs/hep-ph/0702059}{{\tt hep-ph/0702059}}].

\bibitem{Kopp:2007mi}
J.~Kopp, M.~Lindner, and T.~Ota, {\it {Discovery reach for non-standard
  interactions in a neutrino factory}},  {\em Phys.Rev.} {\bf D76} (2007)
  013001, [\href{http://arxiv.org/abs/hep-ph/0702269}{{\tt hep-ph/0702269}}].

\bibitem{Kopp:2007ne}
J.~Kopp, M.~Lindner, T.~Ota, and J.~Sato, {\it {Non-standard neutrino
  interactions in reactor and superbeam experiments}},  {\em Phys.Rev.} {\bf
  D77} (2008) 013007, [\href{http://arxiv.org/abs/0708.0152}{{\tt
  arXiv:0708.0152}}].

\bibitem{Ribeiro:2007ud}
N.~Ribeiro, H.~Minakata, H.~Nunokawa, S.~Uchinami, and R.~Zukanovich-Funchal,
  {\it {Probing Non-Standard Neutrino Interactions with Neutrino Factories}},
  {\em JHEP} {\bf 0712} (2007) 002, [\href{http://arxiv.org/abs/0709.1980}{{\tt
  arXiv:0709.1980}}].

\bibitem{Bandyopadhyay:2007kx}
{\bf ISS Physics Working Group} Collaboration, A.~Bandyopadhyay {\em et~al.},
  {\it {Physics at a future Neutrino Factory and super-beam facility}},  {\em
  Rept.Prog.Phys.} {\bf 72} (2009) 106201,
  [\href{http://arxiv.org/abs/0710.4947}{{\tt arXiv:0710.4947}}].

\bibitem{Ribeiro:2007jq}
N.~C. Ribeiro, H.~Nunokawa, T.~Kajita, S.~Nakayama, P.~Ko, {\em et~al.}, {\it
  {Probing Nonstandard Neutrino Physics by Two Identical Detectors with
  Different Baselines}},  {\em Phys.Rev.} {\bf D77} (2008) 073007,
  [\href{http://arxiv.org/abs/0712.4314}{{\tt arXiv:0712.4314}}].

\bibitem{EstebanPretel:2008qi}
A.~Esteban-Pretel, J.~W. Valle, and P.~Huber, {\it {Can OPERA help in
  constraining neutrino non-standard interactions?}},  {\em Phys.Lett.} {\bf
  B668} (2008) 197--201, [\href{http://arxiv.org/abs/0803.1790}{{\tt
  arXiv:0803.1790}}].

\bibitem{Blennow:2008ym}
M.~Blennow, D.~Meloni, T.~Ohlsson, F.~Terranova, and M.~Westerberg, {\it
  {Non-standard interactions using the OPERA experiment}},  {\em Eur.Phys.J.}
  {\bf C56} (2008) 529--536, [\href{http://arxiv.org/abs/0804.2744}{{\tt
  arXiv:0804.2744}}].

\bibitem{Kopp:2008ds}
J.~Kopp, T.~Ota, and W.~Winter, {\it {Neutrino factory optimization for
  non-standard interactions}},  {\em Phys.Rev.} {\bf D78} (2008) 053007,
  [\href{http://arxiv.org/abs/0804.2261}{{\tt arXiv:0804.2261}}].

\bibitem{Ohlsson:2008gx}
T.~Ohlsson and H.~Zhang, {\it {Non-Standard Interaction Effects at Reactor
  Neutrino Experiments}},  {\em Phys.Lett.} {\bf B671} (2009) 99--104,
  [\href{http://arxiv.org/abs/0809.4835}{{\tt arXiv:0809.4835}}].

\bibitem{Palazzo:2009rb}
A.~Palazzo and J.~Valle, {\it {Confusing non-zero theta(13) with non-standard
  interactions in the solar neutrino sector}},  {\em Phys.Rev.} {\bf D80}
  (2009) 091301, [\href{http://arxiv.org/abs/0909.1535}{{\tt
  arXiv:0909.1535}}].

\bibitem{Blennow:2008eb}
M.~Blennow and T.~Ohlsson, {\it {Approximative two-flavor framework for
  neutrino oscillations with non-standard interactions}},  {\em Phys.Rev.} {\bf
  D78} (2008) 093002, [\href{http://arxiv.org/abs/0805.2301}{{\tt
  arXiv:0805.2301}}].

\bibitem{Davidson:2003ha}
S.~Davidson, C.~Pena-Garay, N.~Rius, and A.~Santamaria, {\it {Present and
  future bounds on nonstandard neutrino interactions}},  {\em JHEP} {\bf 0303}
  (2003) 011, [\href{http://arxiv.org/abs/hep-ph/0302093}{{\tt
  hep-ph/0302093}}].

\bibitem{Biggio:2009kv}
C.~Biggio, M.~Blennow, and E.~Fernandez-Martinez, {\it {Loop bounds on
  non-standard neutrino interactions}},  {\em JHEP} {\bf 0903} (2009) 139,
  [\href{http://arxiv.org/abs/0902.0607}{{\tt arXiv:0902.0607}}].

\bibitem{Apollonio:1999ae}
{\bf CHOOZ} Collaboration, M.~Apollonio {\em et~al.}, {\it {Limits on Neutrino
  Oscillations from the CHOOZ Experiment}},  {\em Phys. Lett.} {\bf B466}
  (1999) 415--430, [\href{http://arxiv.org/abs/hep-ex/9907037}{{\tt
  hep-ex/9907037}}].

\bibitem{Wendell:2010md}
{\bf Super-Kamiokande} Collaboration, R.~Wendell {\em et~al.}, {\it
  {Atmospheric neutrino oscillation analysis with sub-leading effects in
  Super-Kamiokande I, II, and III}},  {\em Phys. Rev.} {\bf D81} (2010) 092004,
  [\href{http://arxiv.org/abs/1002.3471}{{\tt arXiv:1002.3471}}].

\bibitem{Adamson:2011ig}
{\bf The MINOS Collaboration} Collaboration, P.~Adamson {\em et~al.}, {\it
  {Measurement of the neutrino mass splitting and flavor mixing by MINOS}},
  \href{http://arxiv.org/abs/1103.0340}{{\tt arXiv:1103.0340}}.

\bibitem{Adamson:2011fa}
{\bf MINOS} Collaboration, P.~Adamson {\em et~al.}, {\it {First direct
  observation of muon antineutrino disappearance}},
  \href{http://arxiv.org/abs/1104.0344}{{\tt arXiv:1104.0344}}.

\bibitem{Adamson:2010uj}
{\bf The MINOS} Collaboration, P.~Adamson {\em et~al.}, {\it {New constraints
  on muon-neutrino to electron-neutrino transitions in MINOS}},  {\em Phys.
  Rev.} {\bf D82} (2010) 051102, [\href{http://arxiv.org/abs/1006.0996}{{\tt
  arXiv:1006.0996}}].

\bibitem{Kopp:2010qt}
J.~Kopp, P.~A. Machado, and S.~J. Parke, {\it {Interpretation of MINOS data in
  terms of non-standard neutrino interactions}},  {\em Phys.Rev.} {\bf D82}
  (2010) 113002, [\href{http://arxiv.org/abs/1009.0014}{{\tt
  arXiv:1009.0014}}].

\bibitem{Mann:2010jz}
W.~Mann, D.~Cherdack, W.~Musial, and T.~Kafka, {\it {Apparent multiple $\Delta
  m^2_{32}$ in muon anti-neutrino and muon neutrino survival oscillations from
  non-standard interaction matter effect}},  {\em Phys.Rev.} {\bf D82} (2010)
  113010, [\href{http://arxiv.org/abs/1006.5720}{{\tt arXiv:1006.5720}}].

\bibitem{Akhmedov:2010vy}
E.~Akhmedov and T.~Schwetz, {\it {MiniBooNE and LSND data: Non-standard
  neutrino interactions in a (3+1) scheme versus (3+2) oscillations}},  {\em
  JHEP} {\bf 1010} (2010) 115, [\href{http://arxiv.org/abs/1007.4171}{{\tt
  arXiv:1007.4171}}].

\bibitem{Fogli:2008jx}
G.~Fogli, E.~Lisi, A.~Marrone, A.~Palazzo, and A.~Rotunno, {\it {Hints of
  $\theta_{13} > 0$ from global neutrino data analysis}},  {\em Phys.Rev.Lett.}
  {\bf 101} (2008) 141801, [\href{http://arxiv.org/abs/arXiv:0806.2649}{{\tt
  arXiv:0806.2649}}].

\end{thebibliography}\endgroup

\end{document}